\documentclass[%
 aip,pop,
 amsmath,amssymb,
preprint,%
]{revtex4-1}
\usepackage{graphicx}
\usepackage{dcolumn}
\usepackage{bm}
\usepackage{caption}
\usepackage[font=small]{subcaption}
\usepackage{tabularx}

\begin{document}

\title{Ion heating, burnout of the HF field and ion sound generation under the development of a 
modulation instability of an intense Langmuir wave in a plasma}%

\author{A.V.~Kirichok}
\email{sandyrcs@gmail.com}
\author{V.M.~Kuklin}
\author{A.V.~Pryimak}
\affiliation{V.N.~Karazin Kharkiv National University , Institute for High Technologies\\4 Svobody Sq., Kharkiv 61022, Ukraine}

\author{A.G.~Zagorodny}
\affiliation{Bogolyubov Institute for Theoretical Physics\\14-b, Metrolohichna str., Kiev, 03680, Ukraine}

\date{\today}%

\begin{abstract}

The development of one-dimensional parametric instabilities of intense long-wave plasma waves is considered in terms of the so-called hybrid models, when electrons are treated as a fluid and ions are regarded as particles. The analysis is performed for both cases when the average plasma field energy is lower (Zakharov's hybrid model -- ZHM) or greater (Silin's hybrid model -- SHM) than the plasma thermal energy. The efficiency of energy transfer to ions and to ion perturbations under the development of the instability is considered for various values of electron-to-ion mass ratios. The energy of low-frequency (LF) oscillations (ion-sound waves) is found to be much lower than the final ion kinetic energy. We also discuss the influence of the changes in the damping rate of the high-frequency (HF) field on the instability development. Reduced absorption of the HF field leads to the retardation of the HF field burnout within plasma density cavities and to the broadening of the HF spectrum. At the same time, the ion velocity distribution tends to the normal distribution in both ZHM and SHM.


\end{abstract}
\maketitle

\section{Introduction}
Powerful fields that can be excited and are able to propagate in a plasma, i.e. in a nonlinear medium with a wide range of characteristic oscillations, often produce a great variety of instabilities. As a result, the new generally unsteady states come into existence in the medium. The theoretical description of all these successive relaxation processes in the form of cascades of instabilities requires solving very complex problems (in the mathematical sense) with the use of numerical methods and numerical experiments.

Investigating the after-effect of instabilities is important for a variety of applications in plasma physics and plasma electronics. Moreover, the employment of more and more powerful sources of energy leads in particular to the use of plasma in the conventional vacuum high-current electronic devices. The experimental and practical activities in space must also take into account the presence of plasmas.  The very great variety of instabilities and even cascades of instabilities create problems for the implementation of the technologies developed up to now. On the other hand, the abundance of various unstable states provides opportunities for discovering and developing entirely new technologies.

Among the variety of plasma instabilities, the class of parametric instabilities should be especially emphasized. The modulation of velocities and densities of plasma particles by intense external fields results in the development of instabilities similar to those described by Mathieu's or Hill's equations. In the case when the external radiation is in resonance with plasma eigenmodes, such instabilities can be interpreted as decay or modulation ones.
The interest to the parametric instability of intense Langmuir waves, which can be easily excited in the plasma by various sources \cite{Silin.1961, Basov.1964, Dawson.1964, Pashinin.1971, Buts.2006, Fainberg.2000, Kuzelev.1990, Shapiro.1976, Kondratenko.1988} was stipulated, in particular, by new possibilities in heating electrons and ions in a plasma. The correct methods for the description of the parametric instability of long-wave plasma waves were developed in the pioneer works of V.P.~Silin \cite{Silin.1965} and V.E.~Zakharov \cite{Zakharov.1967, Zakharov1.1967, Zakharov.1972}.

The theoretical concepts proposed by V.P.~Silin\cite{Silin.1965} were confirmed by the early numerical experiments on the one-dimensional simulation of the parametric decay of plasma oscillations \cite{Kruer.1970} (see also article\cite{Aliev.1965, Gorbunov.1965}, and review \cite{Silin.1973}).  However, the greatest experimenters' interest was provoked by the mechanism of wave-energy dissipation discovered by V.E.~Zakharov. The analytical studies, laboratory-based experiments, and numerical simulations, performed at an early stage of studying these phenomena \cite{Kruer.1972, Ivanov.1974, Kim.1974}, have confirmed the fact that in some cases, in the course of instability, a significant part of the pump-field energy is transformed into the energy of short-wave Langmuir oscillations \cite{Vyacheslavov.1995, Mcfarland.1997} accompanied with bursts of fast particles \cite{Kruer.1972, Ivanov.1974, Kim.1974, Vyacheslavov.1995, Mcfarland.1997, Andreev.1977, Kovrizhnykh.1977, Buchelnikova.1979, Antipov.1976, Sagdeev.1980, Wong.1984, Cheung.1985, Karfidov.1990,  Zakharov.1989}. The modulation instability of intense Langmuir waves in non-isothermal plasmas also leads to collective ion perturbations, in particular, to the generation of ion-sound waves \cite{Galeev.1975, Sigov.1976, Sigov.1979, Robinson.1999}.

In Zakharov's model \cite{Zakharov.1967} that describes the instability of intense long-wave Langmuir waves in a non-isothermal plasma, namely the modulation instability, results in the excitation of a range of short-wave oscillations. In Silin's model,  a strong Langmuir wave in a cold plasma leads to intense oscillations of the electron velocity with the amplitudes comparable to the wavelengths of the excited modes. In this case, the instability in the general case should be referred as parametric \cite{Silin.1965}. Nevertheless, both these processes are similar \cite{Kuklin.2013, Kuznetsov.1976}. Hence, the term ``modulation instability'' can be applied to  the instability of a strong Langmuir field within the framework of Silin's model.

In the works \cite{Belkin.2013, Kirichok.2014} an attempt was made to compare these models, which have similar physical nature, by the example of one-dimensional description. The choice of the one-dimensional approach, as was noted by J. Dawson \cite{Dawson.1962}, ``often keeps the main features of the processes, but simplifies their description and leads to a fuller understanding of what the important phenomena are'' (see also \cite{Wang.1996}).  Of particular interest is the process of ion heating, so in this paper we use the particle (or finite-sized particle) description for ions because the account of inertial effects can be significant just at the nonlinear stage of the process \cite{Kuklin.1990}.  It was observed \cite{Kuklin.1990, Clark.1992} that simulation in terms of the so-called hybrid model (incorporating one of Zakharov's equations for the HF waves and using particle simulation for ions) demonstrates that fluctuations of ion density are rather significant and favor the development of parametric instability. The non-resonant interaction between ions and HF plasma oscillations, along with ion trapping by the potential wells produced by these oscillations, lead to an instability of density cavities resulting from the modulation instability and produce fast particle groups. 

Clark et al. \cite{Clark.1992} compared the hybrid model with Zakharov's hydrodynamic model. Due to higher level of ion density fluctuations the number of cavities in the hybrid model is found to be significantly greater than in Zakharov's model and their depth is smaller \cite{Henri.2013}. Integral characteristics of the instability after-effects for both models turn out to be essentially identical.

In papers \cite{Belkin.2013, Kirichok.2014} the simulation of one-dimensional ion dynamics was performed in terms of the particle method \cite{Charles.1985, Hockney.1988}. The number of particles used in numerical calculations was $2\cdot 10^4$, which is equivalent to the number of ions, about $(2 \cdot 10^4)^3 \sim 10^{13}$, in the three-dimensional case, in agreement with the conditions of most experiments. Thus, the interaction between modeling particles and plasma oscillations in this simulation is in rather good accordance with the interaction between real particles and plasma waves, naturally with regard to the inherent limitations of the one-dimensional description.  Nevertheless, there is reason to believe that the description of field energy transfer to ions within the framework of the hybrid model represents the real conditions of ion heating by intense Langmuir oscillations in plasmas. Moreover, the one-dimensional description makes it possible to select arbitrary electron-to-ion mass ratios. 

Below, we discuss the efficiency of energy transfer from Langmuir oscillations to ions and ion density perturbations under the development of the modulation instability in both cases of non-isothermic hot and cold one-dimensional plasmas  within the framework of hybrid models and for different values of the electron-to-ion mass ratio. The attention is mainly concentrated on the effect of HF field burnout within density cavities accompanied by energy transfer to the ion component of the plasma.

\section{The hybrid models (HM) of parametric instability}

\subsection{The hybrid model based on Silin's equations (SHM)}

When the intensity of the external electric field is much greater than the specific thermal energy of plasma electrons $W=|E_{0} |^{2} /4\pi \gg n_{0} T_{e} $, it is reasonable to explore the approach presented by V.P.~Silin \cite{Silin1.1973}.

Let consider a one-dimensional plasma system where an intense plasma wave with the wavelength $\lambda _{0} $ and frequency $\omega _{0} $ is excited by an external source. This intense wave will be referred to as the pumping wave.  Since the parametric instability results in the growth of oscillations with rather small wavelength $\lambda \ll \lambda _{0} $, the pumping wave can be suggested to be spatially uniform within the region of interaction:
\begin{equation}
E_{0} =-i(|E_{0} |\exp \{ i\omega _{0} t+i\phi \} -|E_{0} |\exp \{ -i\omega _{0} t-i\phi \} )/2,
\label{eq1}
\end{equation}

\noindent where $|E_{0} |$and $\phi $ are the slowly varying wave amplitude and the phase respectively, $\omega_{0}$ is the external wave frequency, $n_{0}$ and $T_{e}$ are the density and temperature of plasma electrons. Charged particles of plasma oscillate under the action of the electric field and their velocities can be written as $u_{0\alpha } =-\left({e_{\alpha } \left|E_{0} \right| / m_{\alpha } } \omega_{0}\right) \cos \phi =-\omega _{0} b_0 \cos \phi$, where $b_0={e_{\alpha } \left|E_{0} \right| /m_{\alpha }\omega_0^2 }$ is the particle oscillation amplitude.

The equations, governing the nonlinear dynamics of the parametric instability of an intense plasma wave, were derived in the work\cite{Kuklin.2013}. The equations for the HF plasma field spectrum modes $E=\sum _{n}E_{n}  (t)\cdot \exp (ink_{0} x)$ (plasma electrons are treated as fluid and described by hydrodynamic equations) are given by
\begin{multline}
\frac{\partial E_{n} }{\partial t} -i\frac{\omega _{pe}^{2} -\omega _{0} ^{2} }{2\omega _{0} } E_{n} +\theta \frac{{{n^6}}}{{n_M^6}} {E_n}-\frac{4\pi \omega _{pe} \nu _{i,n} }{k_{0} n} J_{1} (a_{n} ) \exp (i\phi )- \\
-i\frac{\omega _{0} }{2en_{0} } \sum _{m}\nu _{i,n-m}   [E_{-m}^{*} J_{2} (a_{n-m} )e^{2i\phi } +E_{m}  J_{0} (a_{n-m} )]=0.
\label{eq2}
\end{multline}
\noindent Here $\omega _{pe} =\sqrt{4\pi e^{2} n_{0} /m_{e} } $ is the background electron plasma frequency, $e$ and $m_{e}$ are the  electron mass and charge, $M$ is the ion mass of an ion, $E_{n} =|E_{n} |\cdot \exp (i\psi _{n} )$ is a slowly varying complex amplitude of the electric field of electron plasma oscillations whose wave-number is $k_{n} =nk_{0}$, $k_{0} ={2\pi /L}$, where $L$ is  the characteristic dimension of the plasma system, $\nu _{i} =\sum _{n}\nu _{i,n}  (t)\cdot \exp (ink_{0} x)$ is the ion charge density, $a_{n} =a\cdot n$, $n,m$ are nonzero integers, and $\pm 1$, e.g. ${a_n} = n{k_0}b = n\left( e{k_0}{E_0}/{m_e} \omega _0^2\right)$. The term ${E_n} \theta  {n^6}/{n_M^6}$ in Eq.~(\ref{eq2}) simulates the HF plasma wave damping by electrons, with $n_M=20$.

The equations of ion motion are given by
\begin{equation}
\frac{d^{2} x_{s} }{dt^{2} } =\frac{e}{M} \sum _{n} \bar{E}_{n}  \exp \{ ik_{0} nx_{s} \},
\label{eq3}
\end{equation}

\noindent and the ion density can be determined from
\begin{equation}
n_{i,n} =\nu _{i,n} /e= \frac{n_{0} k_{0} }{2\pi } \int _{-\pi /k_{0} }^{\pi /k_{0} }\exp [-ink_{0}  x_{s} (x_{0} ,t)] dx_{s0}.
\label{eq4}
\end{equation}

The slowly varying electric field strength $\bar{E}_{n} $, acting on the ions, is equal to
\begin{multline}
\bar{E}_{n} =-\frac{4\pi i}{k_{0} n} \nu _{i,n} [1-J_{0}^{2} (a_{n} )+\frac{2}{3} J_{2} ^{2} (a_{n} )]+\\
+\frac{1}{2} J_{1} (a_{n} )[E_{n}  e^{-i\phi } -E_{-n}^{*}  e^{i\phi } ]-
-\frac{ink_{0} }{16\pi en_{0} } J_{0} (a_{n} )\sum _{m}E_{n-m}  E_{-m}^{*}  -\\
-\frac{ik_{0} J_{2} (a_{n} )}{16\pi en_{0} } \sum _{m}(n-m)[E_{n-m}  E_{m}  e^{-2i\phi } +E_{m-n}^{*}  E_{-m}^{*}   e^{2i\phi } ],
\label{eq5}
\end{multline}

The equation for the uniform component of the electric field $E_{0} =\left|E_{0} \right|\exp (i\phi )$ can be written as
\begin{equation}
\frac{\partial E_{0} }{\partial t} =-\frac{\omega _{0} }{2en_{0} } \sum _{m}\nu _{i,-m}   [E_{-m}^{*}  J_{2} (a_{m} )e^{2i\phi } +E_{m}  J_{0} (a_{m} )].
\label{eq6}
\end{equation}

Note, that the values with subscripts with unlike signs are independent. In Eqs.~(\ref{eq2})--(\ref{eq6}), we have used the formula \cite{Dwight.1961}
\begin{equation}
\exp \{ ia \sin\Phi \} =\sum _{m=-\infty }^{\infty }J_{m}  (a) \exp \{ im\Phi \},
\label{eq7}
\end{equation}

\noindent where $J_{m}(x)$ is the Bessel function.

The normalized frequency shift $\Delta =(\omega _{pe}^{2} -\omega _{0} ^{2} )/2\delta \omega _{pe} $ reaches the value of $({m_{e}/2M} )^{1/3} J_{1} ^{2/3} (a_{n} )$ for  the mode with the maximum growth rate of the parametric instability $\delta $ (see \cite{Silin1.1973})
\begin{equation}
\delta /\omega _{pe} =\frac{i}{\sqrt[{3}]{2} } \left(\frac{m_{e} }{M} \right)^{1/3} J_{1} ^{2/3} (a_{n} ).
\label{eq8}
\end{equation}

\subsection{The hybrid model based on Zakharov's equations (ZHM)}

As is shown in \cite{Kuklin.2013}, Eqs.~(\ref{eq2})--(\ref{eq6}) reproduce the equations obtained in \cite{Kuznetsov.1976} after following substitutions: $(\omega _{pe}^{2} -\omega _{0} ^{2} )/2\omega _{0} \to (\omega _{pe}^{2} -\omega _{0} ^{2} +k^{2} _{0} n^{2} v_{Te}^{2} )/2\omega _{0} $ and $E_{0} \to -iE_{0} $, $E_{0}^{*} \to iE_{0}^{*} $ under the condition $a_{n} \ll 1$, that means that $J_{1} (a_{n} )\approx a_{n} /2$, $J_{0} (a_{n} )\approx 1$, $J_{2} (a_{n} )\approx a_{n}^{2} /8$
\begin{equation}
\frac{\partial E_{n} }{\partial t} -i\frac{\omega _{pe}^{2} -\omega _{0} ^{2} +k^{2} _{0} n^{2} v^{2} _{Te} }{2\omega _{0} } E_{n}+\theta \frac{{{n^6}}}{{n_M^6}} {E_n} -i\frac{\omega _{0} }{2n_{0} }  \left( n_{i,n} E_{0} +\sum _{m\ne 0}n_{i,n-m} E_{m}  \right) =0.
\label{eq9}
\end{equation}

\noindent Note that the damping of the HF modes (the term $\propto \theta$) was introduced in Eq.~(\ref{eq9}) phenomenologically, because we are only interested in how the changes in the absorption level influence the ion heating. More precise analysis of the effect of kinetic processes on the Landau damping and on the saturation of parametric instabilities in an electromagnetically driven plasma can be found in papers \cite{Sanbonmatsu.2000, Sanbonmatsu1.2000}.

In this case, the slowly varying electric field amplitude takes the form 
\begin{equation}
\bar{E}_{n} =-\frac{ik_{0} ne}{4m\omega _{p} ^{2} } \left(E_{n} E_{0}^{*} +E_{0} E_{-n}^{*} +\sum _{m\ne 0,n}E_{n-m} E_{-m}^{*}  \right),
\label{eq10}
\end{equation}

\noindent that makes it possible to describe ions as particles within the context of of Eqs.~(\ref{eq3})--(\ref{eq4}). The pump wave amplitude $E_{0}$ is governed by the equation
\begin{equation}
\frac{\partial E_{0} }{\partial t} -i\frac{\omega _{0} }{2n_{0} }  \sum _{m}n_{i,-m} E_{m}  =0.
\label{eq11}
\end{equation}

In this case, the growth rate of the parametric instability normalized to the plasma frequency is given by \cite{Kuznetsov.1976}
\begin{equation}
\delta /\omega _{pe} =\left( \frac{|E_{0} |^{2} }{8\pi n_{0} T_{e} } \frac{m_{e} }{M} \right)^{1/2} =\left(\frac{W}{2 n_{0} T_{e} } \frac{m_{e} }{M} \right)^{1/2}.
\label{eq12}
\end{equation}

\section{Statement of the problem and the initial conditions}

The purpose of this paper is to clarify the efficiency of the energy transfer to ions and ion density perturbations  in the course of development of modulation instability for the cases of both non-isothermal hot and cold plasmas in terms of the hybrid models. Both SHM and ZHM were considered for two cases of light and heavy ions.The parameters of the simulation are presented in Table 1.
It is also interesting to elucidate the effect of HF spectrum damping and subsequent burnout of the Langmuir field within density cavities on the energy transfer to plasma ions.

\begin{table*}[b]  \footnotesize
\caption{Simulation parameters for the hybrid models}
\begin{ruledtabular}
\begin{tabularx}{\textwidth}{ccc}
Model & Light ions  & Heavy ions \\
 &${M}/{m_{e}}  =2\cdot 10^{3}$&$m_{e}/M= 8\cdot 10^{-6} $\\
 \hline \noalign{\smallskip}
 SHM   &  $(m_e/M) (\omega ^{2} _{p} /\delta ^{2})=0.43$
 & $(m_e/M) (\omega ^{2} _{p} /\delta ^{2})=0.1$
 \\
 & ${\delta /}{\omega _{0} } =0.44 \cdot ({m_{e} }/{M} )^{1/3} =0.034$ & $\delta /{\omega _{0} } =0.44\cdot ({m_{e} }/{M} )^{1/3} =0.0088$
 \\
 &${\omega _{0} }/{\delta } \approx {\omega _{pe} }/{\delta } =29.4$ & $\omega _{0} /\delta  \approx {\omega _{pe} }/{\delta } =113.6$ \\ \hline \noalign{\smallskip}
ZHM & $(m_e/M) (\omega ^{2} _{p} /\delta ^{2})=2{n_{0} T_{e} }/{W} =20$&
  $(m_e/M) (\omega ^{2} _{p} /\delta ^{2})=2{n_{0} T_{e} }/{W} =20$
\\
   & $\omega _{0}/\delta  =2\left(n_{0} T_{e}/W \right)^{1/2} (M/m_{e})^{1/2} =282.6$
 &${\omega _{0} }/{\delta } =2(n_{0} T_{e}/W)^{1/2} (M/m_{e})^{1/2} =2234.4$ \\
 & ${\delta }/{\omega _{0} } ={\delta }/{\omega _{pe} } = 3.5 \cdot 10^{-3} $&${\delta }/{\omega _{0} }={\delta }/{\omega _{pe} } =4.5\cdot 10^{-4} $
\end{tabularx}
\end{ruledtabular}
\label{Tab1}
\end{table*}

Below we employ, unless otherwise specified in the text, the following initial conditions and parameters. The number of particles simulating the dynamics of ions is $0<s\le~S=20000$. The particles are distributed uniformly over the interval $-1/2<\xi <1/2$, $\xi =k_{0} x/2\pi$, initial ion velocities are defined as $d\xi _{s} /d\tau |_{\tau =0} =v_{s} |_{\tau =0} =0$, the number of spectrum modes is $-N<n<N$, $N=S/100$. The initial normalized amplitude of the pumping wave is $a_{0} (0)=ek_{0} E_{0} (0)/m_{e} \omega _{pe}^{2} =0.06$,. The initial amplitudes of HF plasma oscillations are defined by the expression $e_{n} |_{\tau =0} =e_{n0} =(2+g_n) \cdot 10^{-3} $ for the Silin model and by the expression $e_{n} |_{\tau =0} =e_{n0} =(0.5+g_n)\cdot 10^{-4} $ for the Zakharov model, where $g_n\in \left[0;1\right]$ is a random value, $ek_{0} E_{n} /m_{e} \omega_{pe}^{2} =e_{n}  \exp (i\psi_{n} )$. The initial phases of spectral modes $\psi_{n} |_{\tau =0} $ are also randomly distributed in the interval $0\div 2\pi $. The ion density fluctuations $n_{ni} $ and slowly varying electric field $\bar{E}_{n} $ are described by the dimensionless representations
$$
M_{n} =M_{nr} +iM_{ni} =n_{ni} \omega _{pe} /n_{0} \delta =\left(\omega _{pe}/{\delta } \right) \int _{-\pi /k_{0} }^{\pi /k_{0} }\exp (2\pi n \xi _{s} ) d\xi _{s0}  
$$

\noindent and
$$ek_{0} \bar{E}_{n} /m_{e} \omega _{pe}^{2} =E_{nr} +iE_{ni}.$$

The development of the instability was considered in terms of hybrid models Eqs.~(\ref{eq2})--(\ref{eq8}) and Eqs.~(\ref{eq9})--(\ref{eq12}) in our previous papers \cite{Belkin.2013, Kirichok.2014}. Here we give some results. The rate of damping of HF modes governs the rate of the field energy burnout in density caverns,  from where the HF field has forced out charged particles  The main part of the instability energy is initially concentrated in the HF Langmuir oscillations in parallel with the formation  of the LF spectrum of density perturbations.  Then the energy of the HF spectrum is transferred mainly to electrons. Thus, the shaped density cavities collapse, the trajectories of ions intercross, ion density perturbations become smoother and their characteristic scale growths with time. The relationship between ionic perturbations and the HF field is weakened and the instability is saturated. The amplitude of the main wave stabilizes after several oscillations at rather low level. The bulk energy is now contained in the perturbations of the electron component of the plasma. Some small portion of the initial energy transforms into the kinetic energy of ions. The estimate of the energy density transmitted to ions $E_{kin}$ can be obtained from the expression

\begin{equation}
\frac{E_{kin}}{W_0} \approx 0.27 \cdot I \cdot \frac{M}{m} \cdot \frac{\delta ^2}{\omega _{pe}^2},
\label{eq13}
\end{equation}

\noindent where $W_0$ is the initial energy density of the intense Langmuir wave, $I=\sum_s {(d\xi_s /d\tau)^2}$ is normalized ion kinetic energy and $\delta$ is the rate of the linear instability. The portion of energy  transferred from the intense Langmuir wave to ions is determined by the ratio $W_0/n_0 T_e$ for the case of non-isothermic plasma (ZHM) and by the ratio $(m/M)^{1/3}$ for the case of cold plasma (SHM).

Below we consider more closely the nature of the energy redistribution with time and especially the process of energy transfer to the LF perturbations.  We also discuss the specific features of the excitation of LF ion-sound waves in both non-isothermic and cold plasmas. More attention will be focused on the role of absorption of HF spectrum that is responsible for the burnout of the HF field in the density cavities. We investigate the effect of this process on the excitation of the LF spectrum and most importantly on the kind of ion velocity distribution function and on the proportion of the total energy transferred to ions.

\section{The results of numerical simulation}

\begin{figure}[t] \centering
	\begin{subfigure}[b]{0.47\textwidth}
		\includegraphics[width=\textwidth]{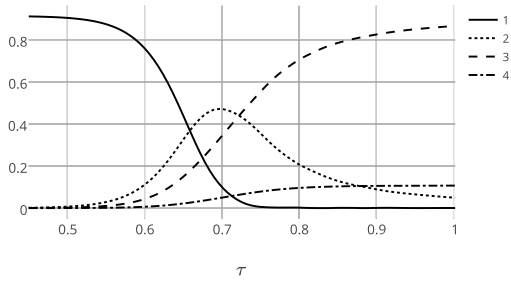}
		\caption{ZHM, light ions}
	\end{subfigure}
	\begin{subfigure}[b]{0.47\textwidth}
		\includegraphics[width=\textwidth]{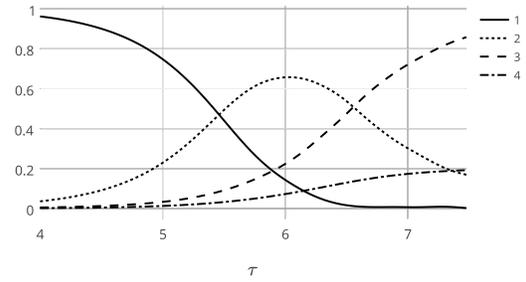}
		\caption{SHM, light ions}
	\end{subfigure}

	\begin{subfigure}[b]{0.47\textwidth}
		\includegraphics[width=\textwidth]{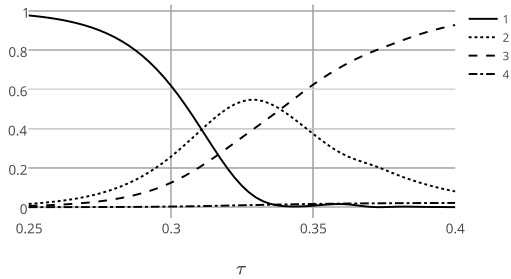}
		\caption{ZHM, heavy ions}
	\end{subfigure}
	\begin{subfigure}[b]{0.47\textwidth}
		\includegraphics[width=\textwidth]{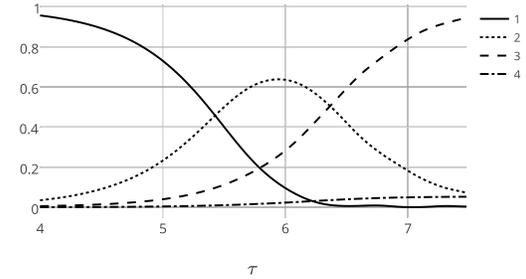}
		\caption{SHM, heavy ions}
	\end{subfigure}
\caption{Time evolution of relative values of:  energy of the main Langmuir wave (1),  energy of low-scale plasma wave spectrum (2), energy transferred to electrons (3) and ions (4).}
\label{fig1}
\end{figure}

Figure \ref{fig1} shows the energy redistribution between the main Langmuir wave, the small-scale plasma wave spectrum and plasma electrons and ions for the following values of parameters determining the damping rate of the HF spectrum: $n_M=20$, $\Theta=0.005$.

The analysis of the numerical simulation results shows that the energy of intense long-wave Langmuir waves is first transferred to short-wave Langmuir oscillations. Just at this stage the cavities of plasma density, filled with HF plasma oscillations, are formed. After that, the HF field burns out due to the damping on electrons that is included in the hybrid models phenomenologically. The energy of the HF field therewith converts into  the energy of plasma electrons. Under these conditions, the cavities collapse and thus excite LF waves, the ion trajectories intercross, and the energy of both collapsed caverns and LF spectrum is transferred to ions.

\begin{figure}[t] \centering
	\begin{subfigure}[b]{0.47\textwidth}
		\includegraphics[width=\textwidth]{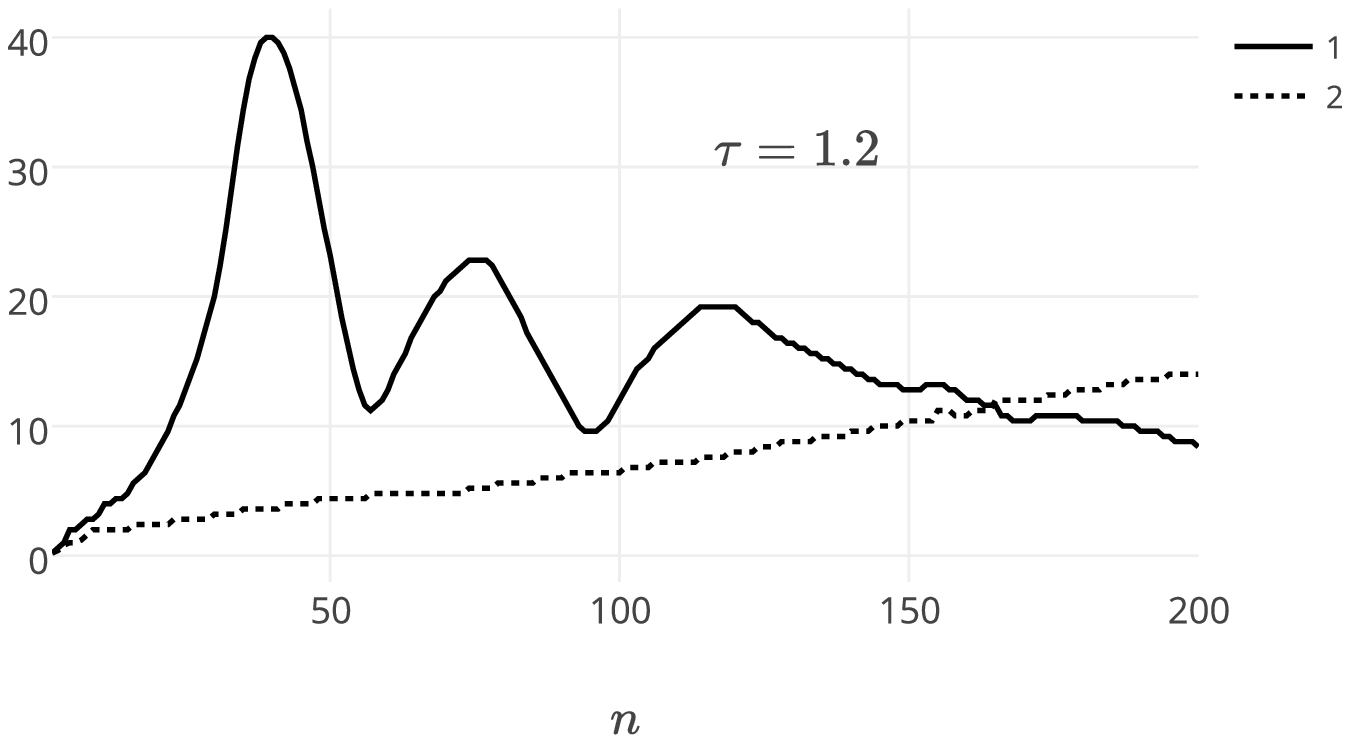}
		\caption{ZHM, light ions}
	\end{subfigure}
	\begin{subfigure}[b]{0.47\textwidth}
		\includegraphics[width=\textwidth]{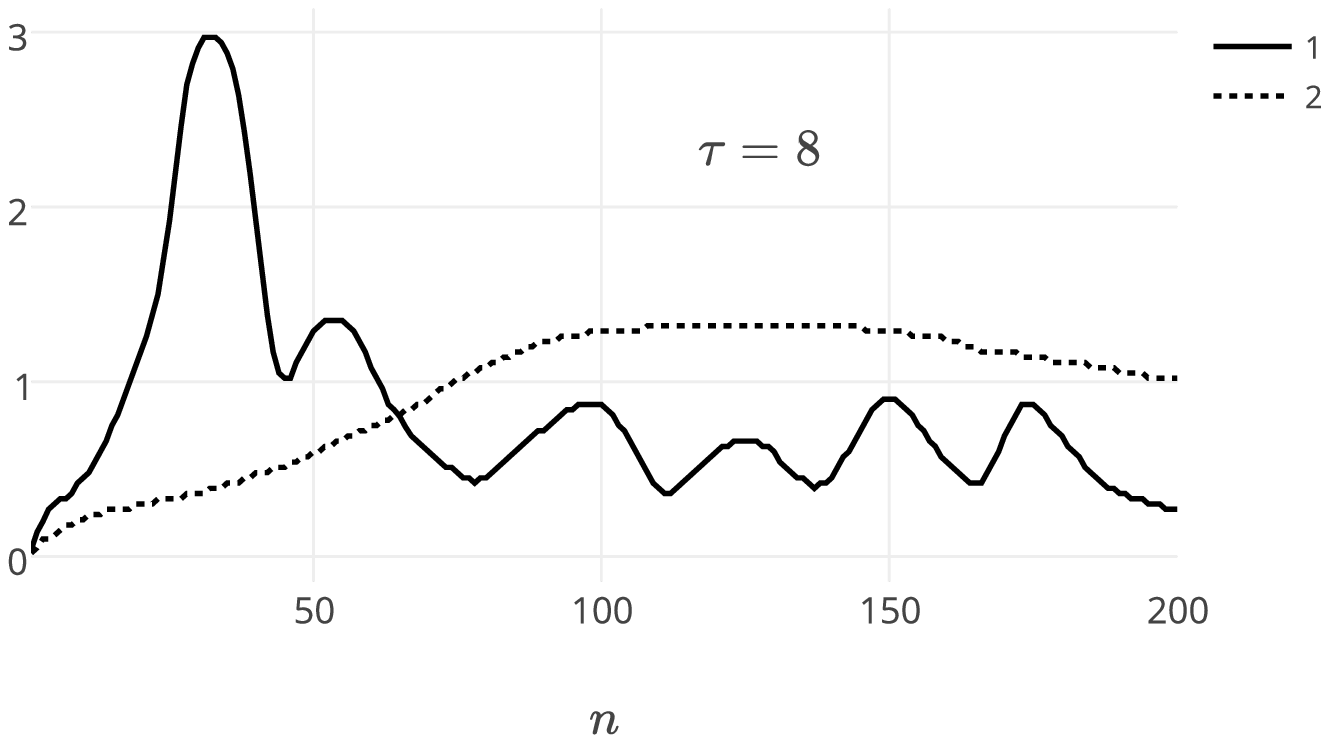}
		\caption{SHM, light ions}
	\end{subfigure}

	\begin{subfigure}[b]{0.47\textwidth}
		\includegraphics[width=\textwidth]{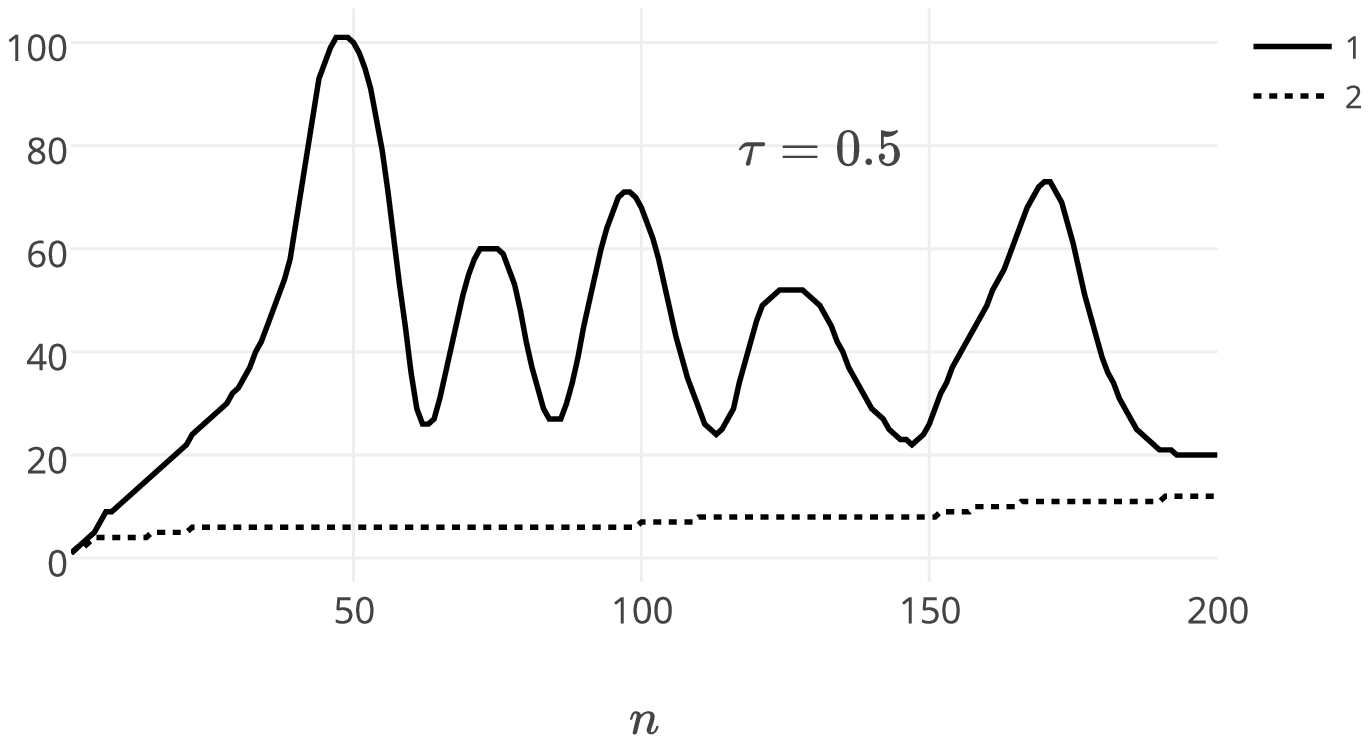}
		\caption{ZHM, heavy ions}
	\end{subfigure}
	\begin{subfigure}[b]{0.47\textwidth}
		\includegraphics[width=\textwidth]{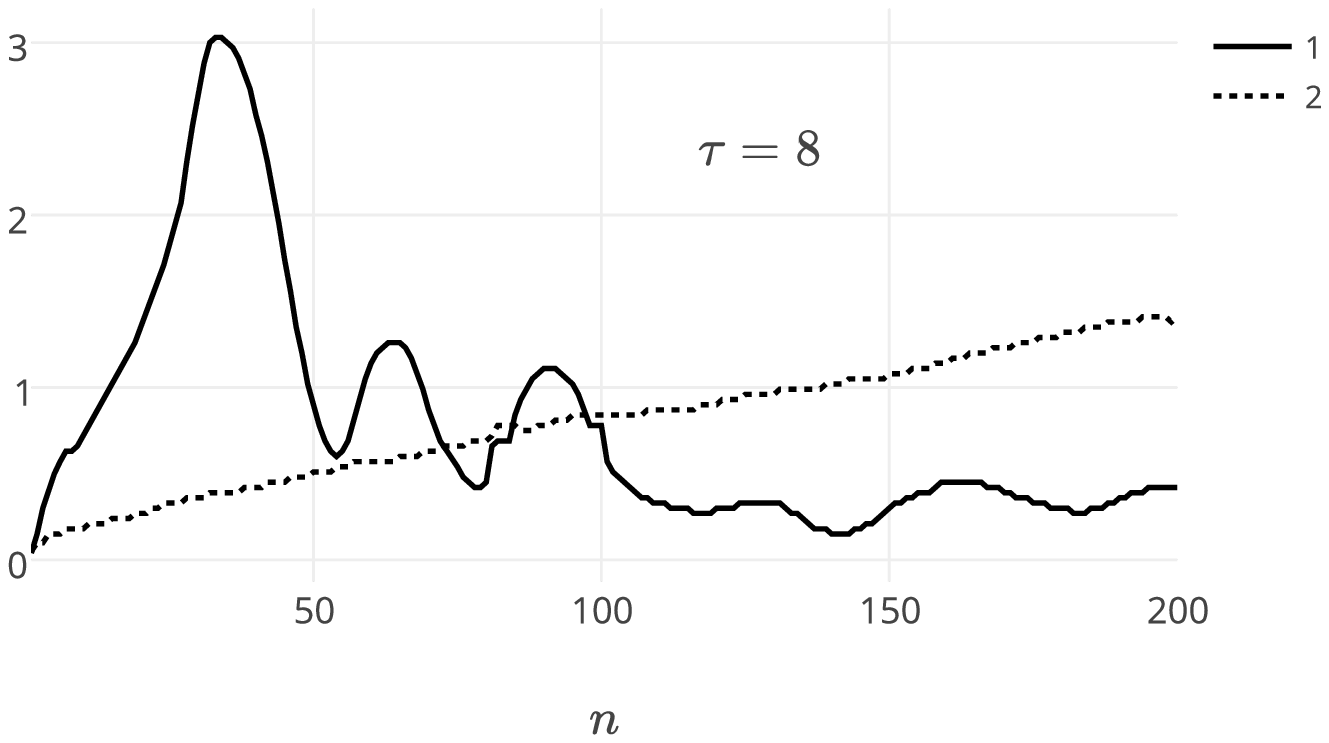}
		\caption{SHM, heavy ions}
	\end{subfigure}
\caption{Dependence of the amplitude of the LF modes $M_n$ (1) and the frequency $d\Phi_n/d\tau$ (2) on the wave-number at the stage of developed instability.}
\label{fig2}
\end{figure}

The root-mean-square velocity of ions, $\sigma (v) = \sqrt {\sum_s {v_s^2} /S}$, at the final stage of the numerical simulation is equal to $\sigma (v) = 0.015$ for the case of light ions and  $\sigma (v) = 0.006$ for heavy ions in ZHM and, respectively, to $\sigma (v) = 0.002$ for light ions and $\sigma (v) = 0.0005$ for heavy ions in SHM.
The total kinetic energy of ions in assumed units $I = {\sum_s {(d\xi _s/d\tau) ^2}}$ is equal to $4.689$ for the case of light ions and  $0.808$ for heavy ions in ZHM and $\sigma (v) = 0.086$ for light ions and $0.005$ for heavy ions in SHM. The variations in the values of the total energy are caused by different linear growth rates  in the two models under consideration, and by different ion masses in the simulation of light and heavy ions. The final ion velocity distribution can be fitted by the normal curve with the use of the values of {\it rms} velocity. The particles outside the normal distribution (mainly in the so-called "tails") possess 13.8\% of the total energy for light ions and 9,2\% for heavy ions in ZHM model and much more in SHM: 25,6\% for heavy ions and 13\% for light ions, respectively. It means that in the case of instability of the intense wave in a cold plasma, a significantly greater proportion of fast particles should be expected.

\begin{figure}[t] \centering
	\begin{subfigure}[b]{0.47\textwidth}
		\includegraphics[width=\textwidth]{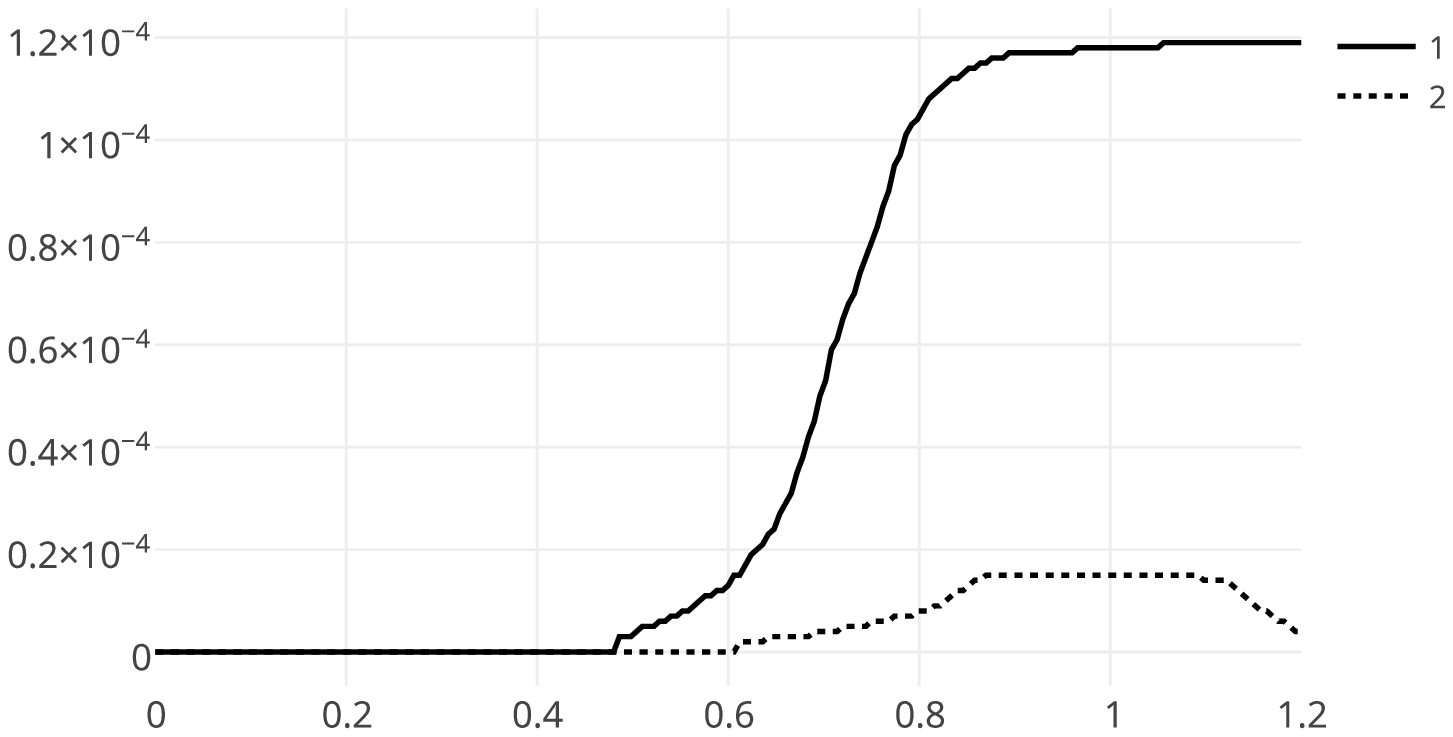}
		\caption{ZHM, light ions}
	\end{subfigure}
	\begin{subfigure}[b]{0.47\textwidth}
		\includegraphics[width=\textwidth]{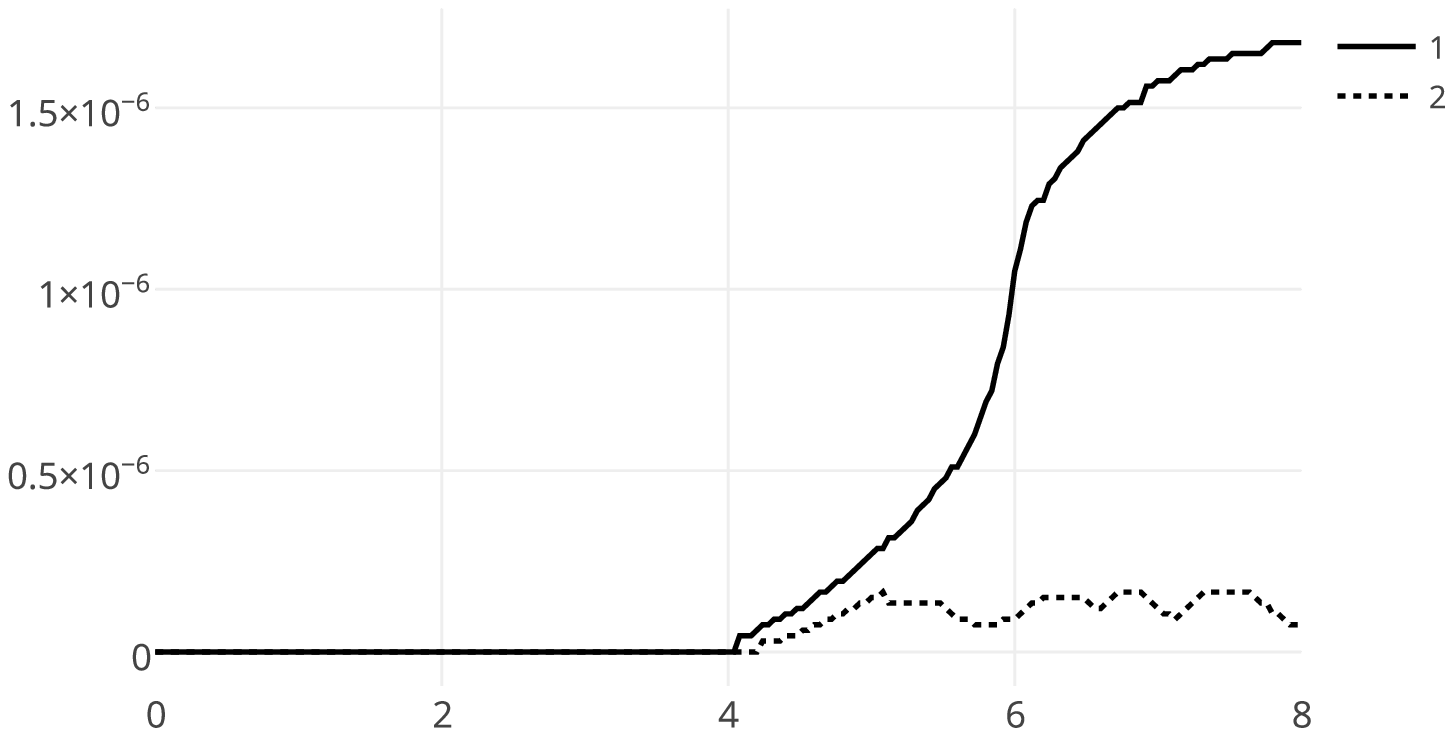}
		\caption{SHM, light ions}
	\end{subfigure}

	\begin{subfigure}[b]{0.47\textwidth}
		\includegraphics[width=\textwidth]{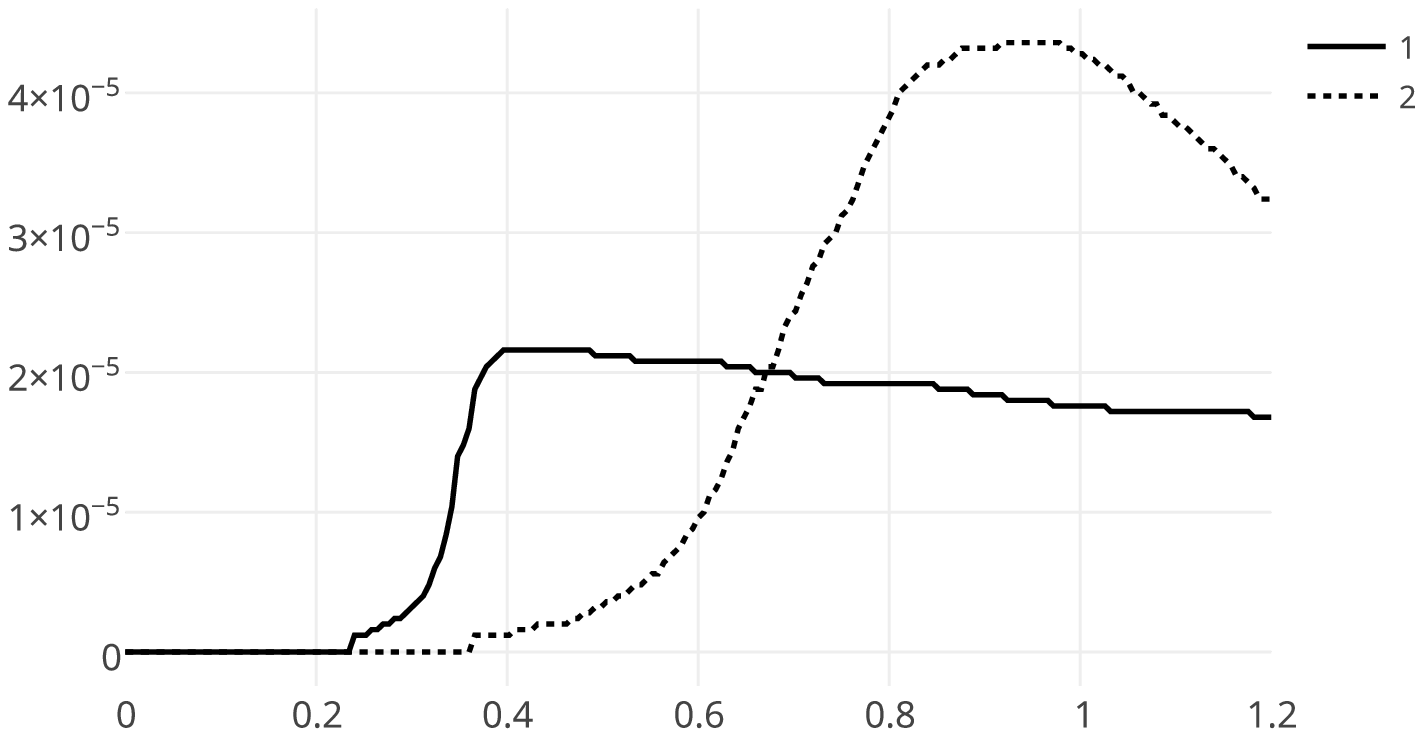}
		\caption{ZHM, heavy ions}
	\end{subfigure}
	\begin{subfigure}[b]{0.47\textwidth}
		\includegraphics[width=\textwidth]{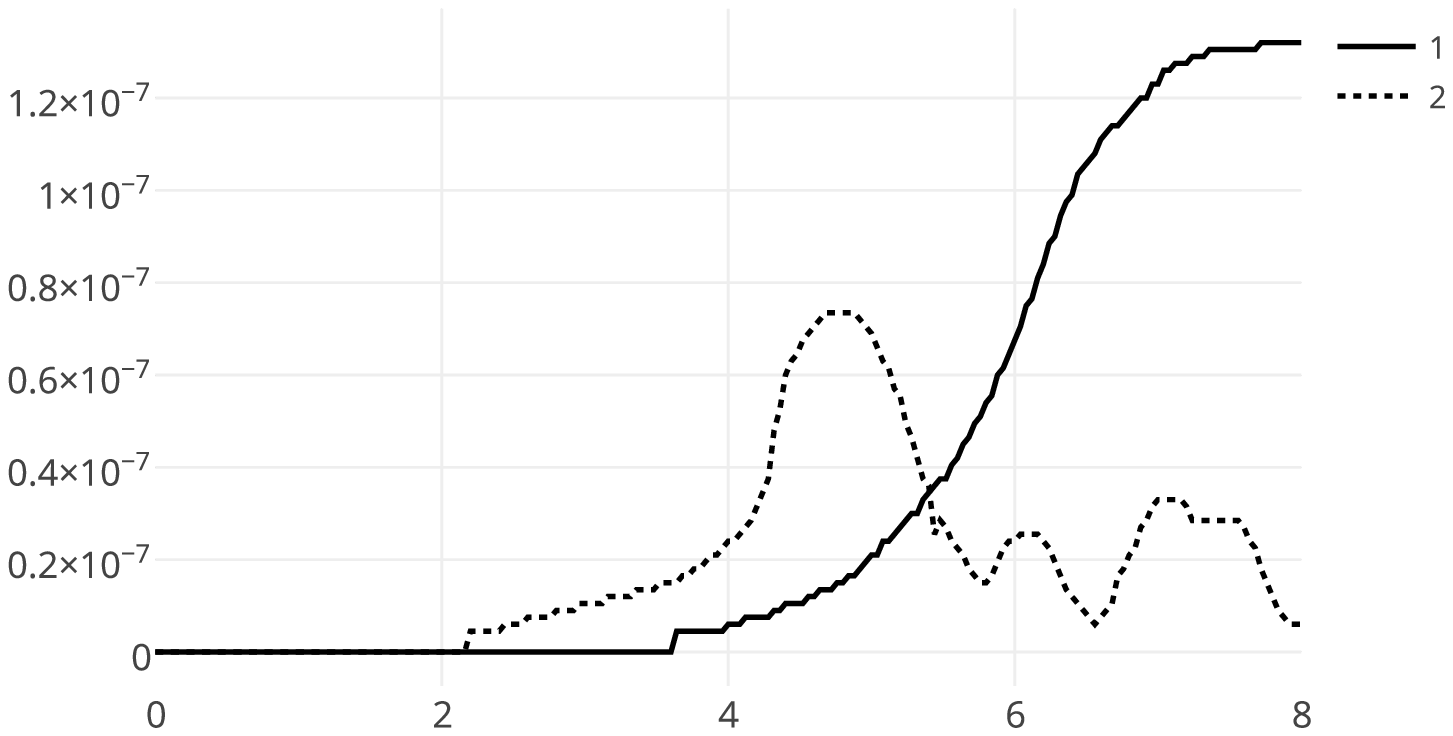}
		\caption{SHM, heavy ions}
	\end{subfigure}
\caption{Evolution of the ion kinetic energy (1) and LF field energy multiplied by factor 70 (2) with time.}
\label{fig3}
\end{figure}

We are interested  not only in the ion kinetic energy distribution, but also in the collective excitation of ion oscillation (see Fig.~\ref{fig2}), hence  we define the frequency of the mode with  the wave vector $n k_0$ associated with these oscillations, i.e.,
\begin{equation}
\frac{{d{\Phi _n}}}{{d\tau }} =  - {{\frac{d}{{d\tau }}\left( {\frac{{{M_{nr}}}}{{\sqrt {M_{nr}^2 + M_{ni}^2} }}} \right)} \mathord{\left/
 {\vphantom {{\frac{d}{{d\tau }}\left( {\frac{{{M_{nr}}}}{{\sqrt {M_{nr}^2 + M_{ni}^2} }}} \right)} {\left( {\frac{{{M_{ni}}}}{{\sqrt {M_{nr}^2 + M_{ni}^2} }}} \right)}}} \right.
 \kern-\nulldelimiterspace} {\left( {\frac{{{M_{ni}}}}{{\sqrt {M_{nr}^2 + M_{ni}^2} }}} \right)}},
 \label{eq14}
\end{equation}
\noindent where the phases of LF modes can be found from the expression
$${M_n} = {M_{nr}} + i{M_{ni}} = \sqrt {M_{nr}^2 + M_{ni}^2}  \cdot \exp (i{\Phi _n}).$$ It should be noticed that the intensity of the LF spectrum in the case of a non-isothermal plasma (ZHM) is quite high in a wide range of wave numbers, that corresponds to the spectrum of ion sound after the destruction of density cavities detected in numerical experiments \cite{Sigov.1979}. In a cold plasma (SHM), in contrast,  the  long-wave oscillations dominate in the spectrum.

\begin{figure}[t] \centering
	\begin{subfigure}[b]{0.47\textwidth}
		\includegraphics[width=\textwidth]{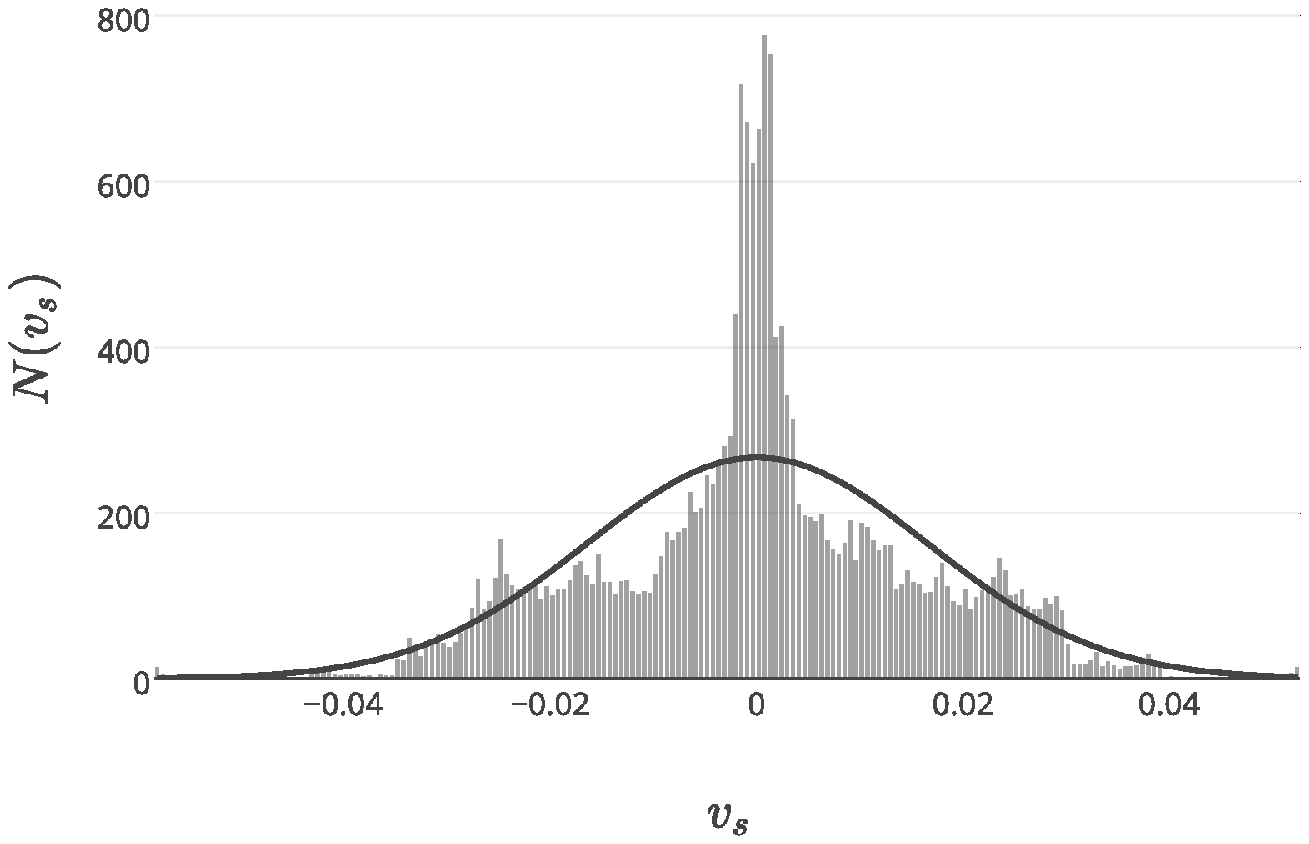}
		\caption{ZHM, $\Theta = 0.05$}
	\end{subfigure}
	\begin{subfigure}[b]{0.47\textwidth}
		\includegraphics[width=\textwidth]{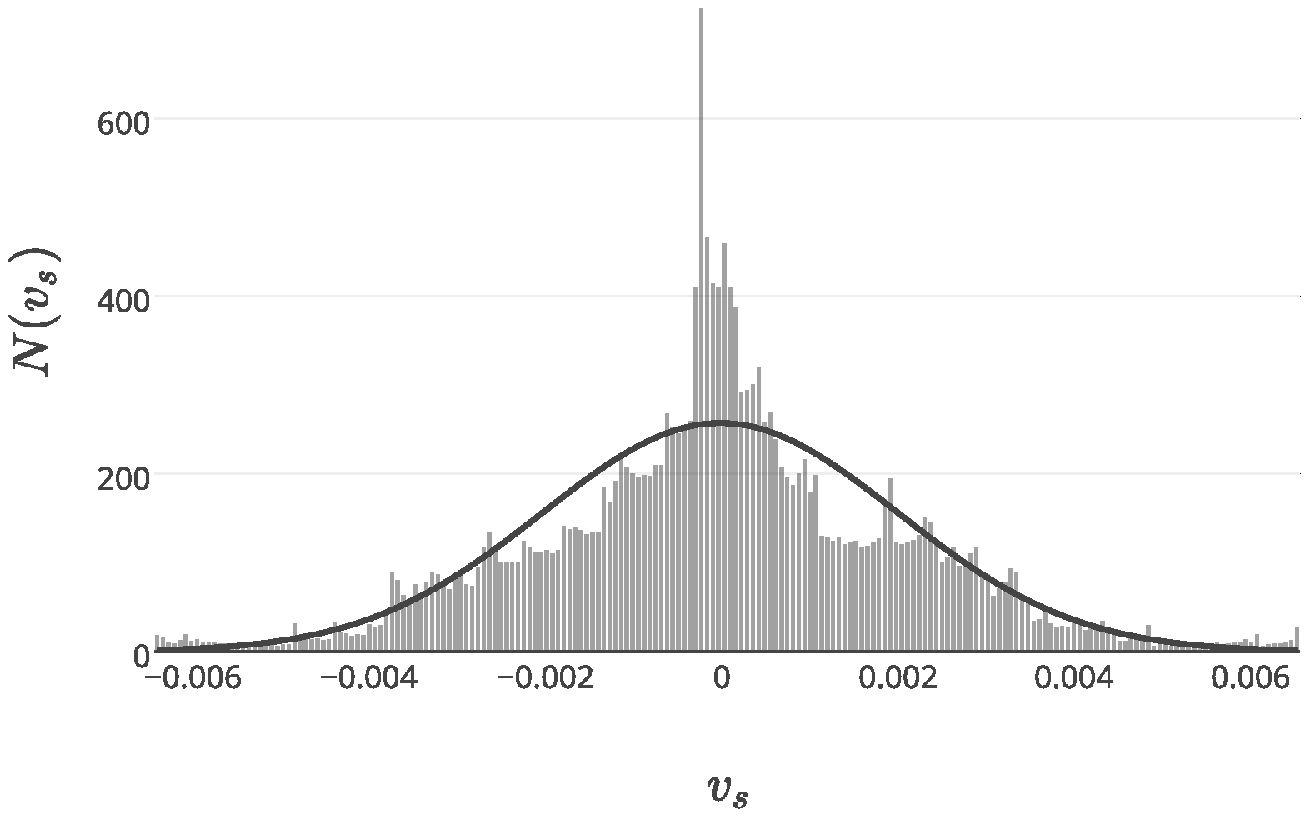}
		\caption{SHM, $\Theta = 0.05$}
	\end{subfigure}

	\begin{subfigure}[b]{0.47\textwidth}
		\includegraphics[width=\textwidth]{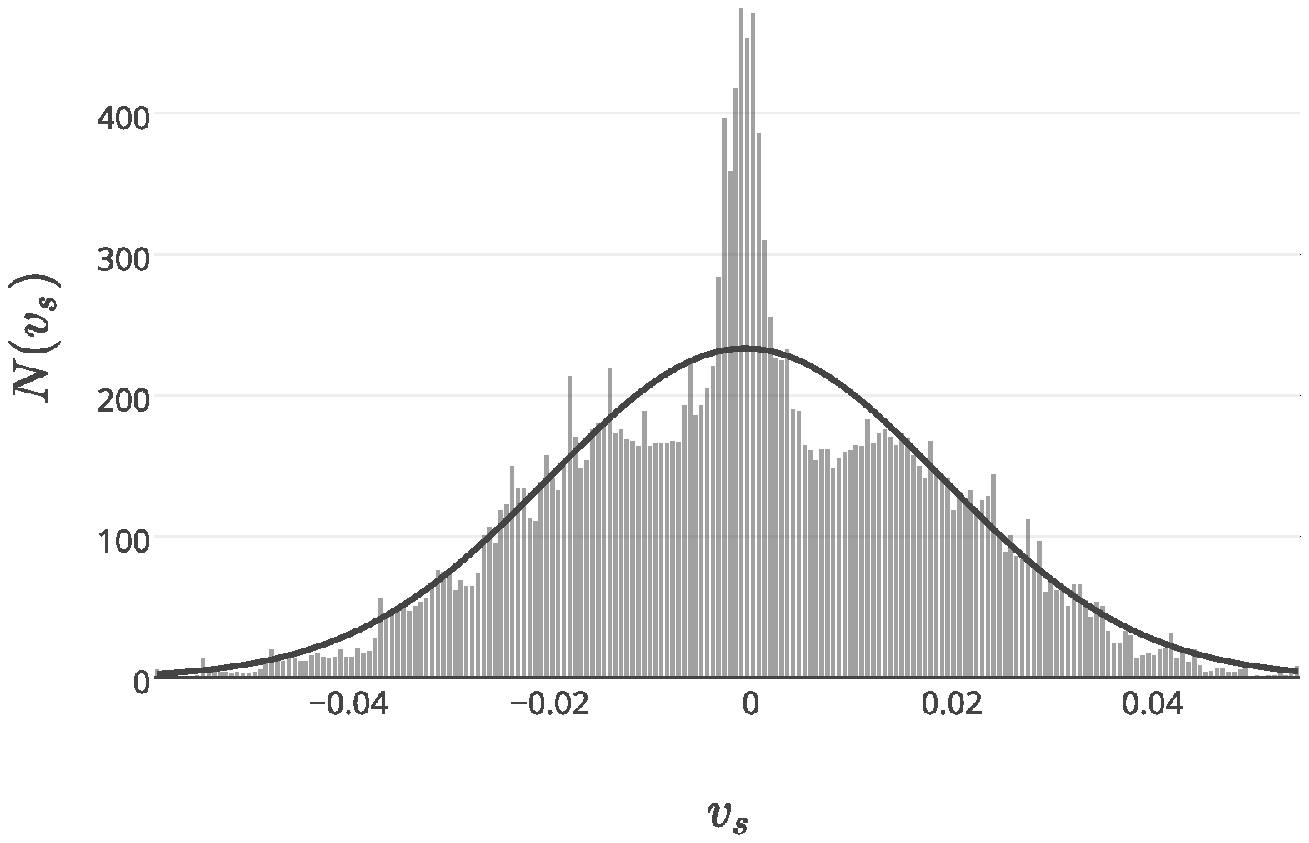}
		\caption{ZHM, $\Theta = 0.015$}
	\end{subfigure}
	\begin{subfigure}[b]{0.47\textwidth}
		\includegraphics[width=\textwidth]{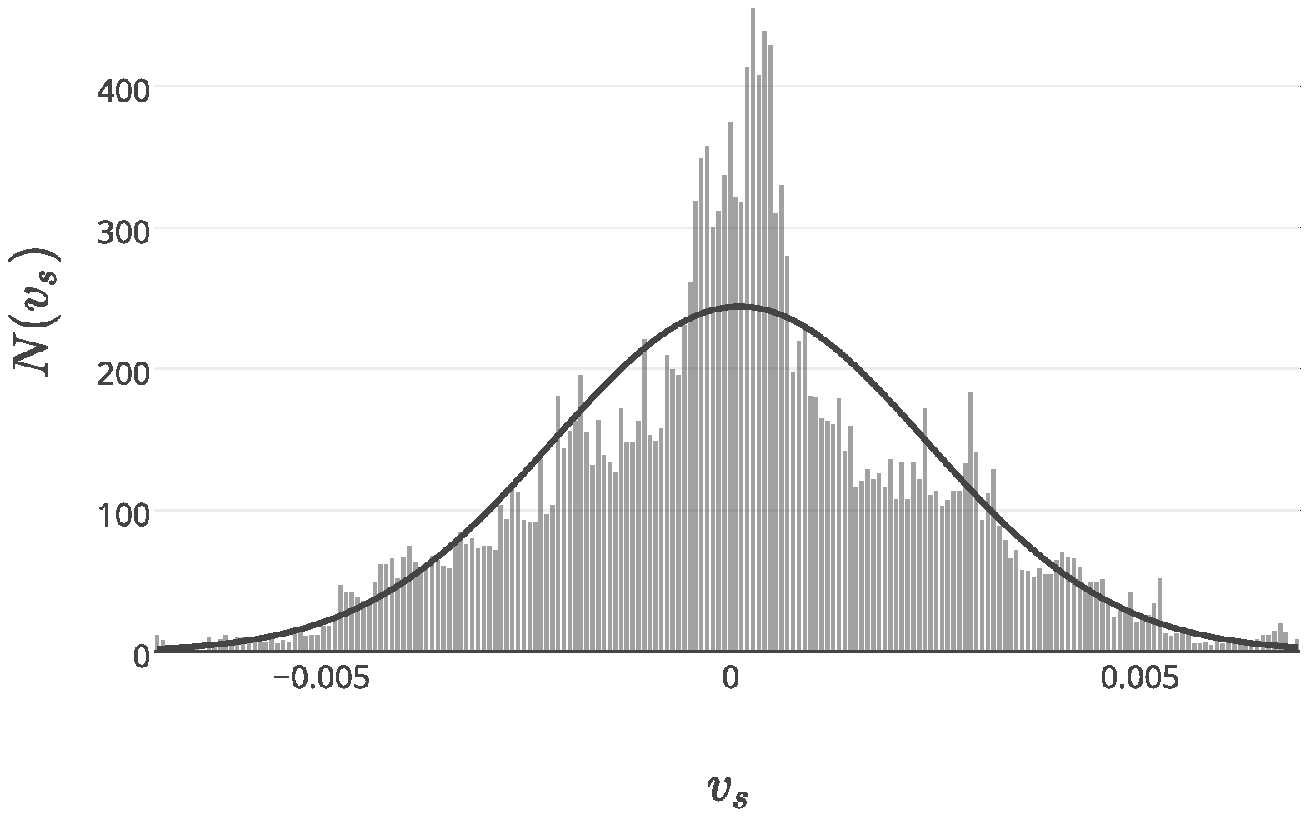}
		\caption{SHM, $\Theta = 0.015$}
	\end{subfigure}

    \begin{subfigure}[b]{0.47\textwidth}
		\includegraphics[width=\textwidth]{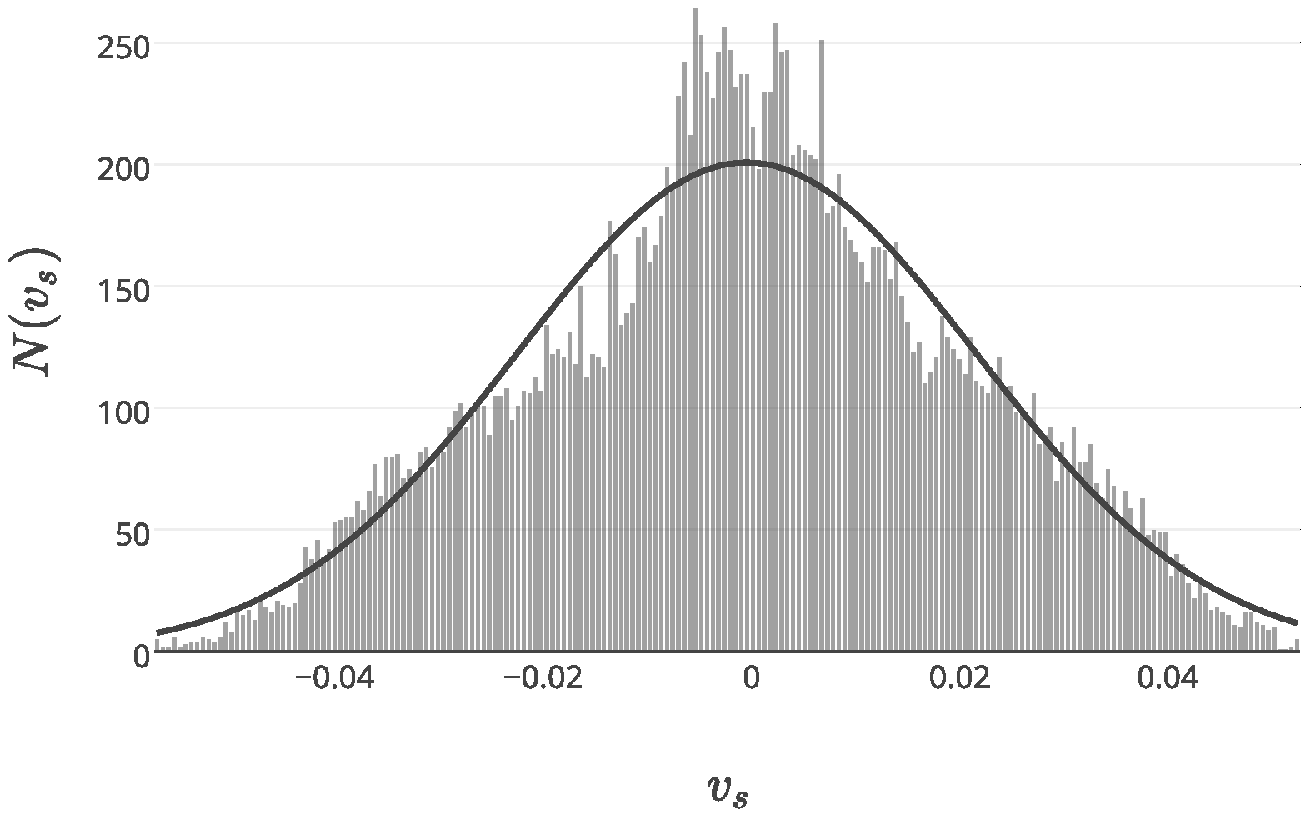}
		\caption{ZHM, $\Theta = 0.001$}
	\end{subfigure}
	\begin{subfigure}[b]{0.47\textwidth}
		\includegraphics[width=\textwidth]{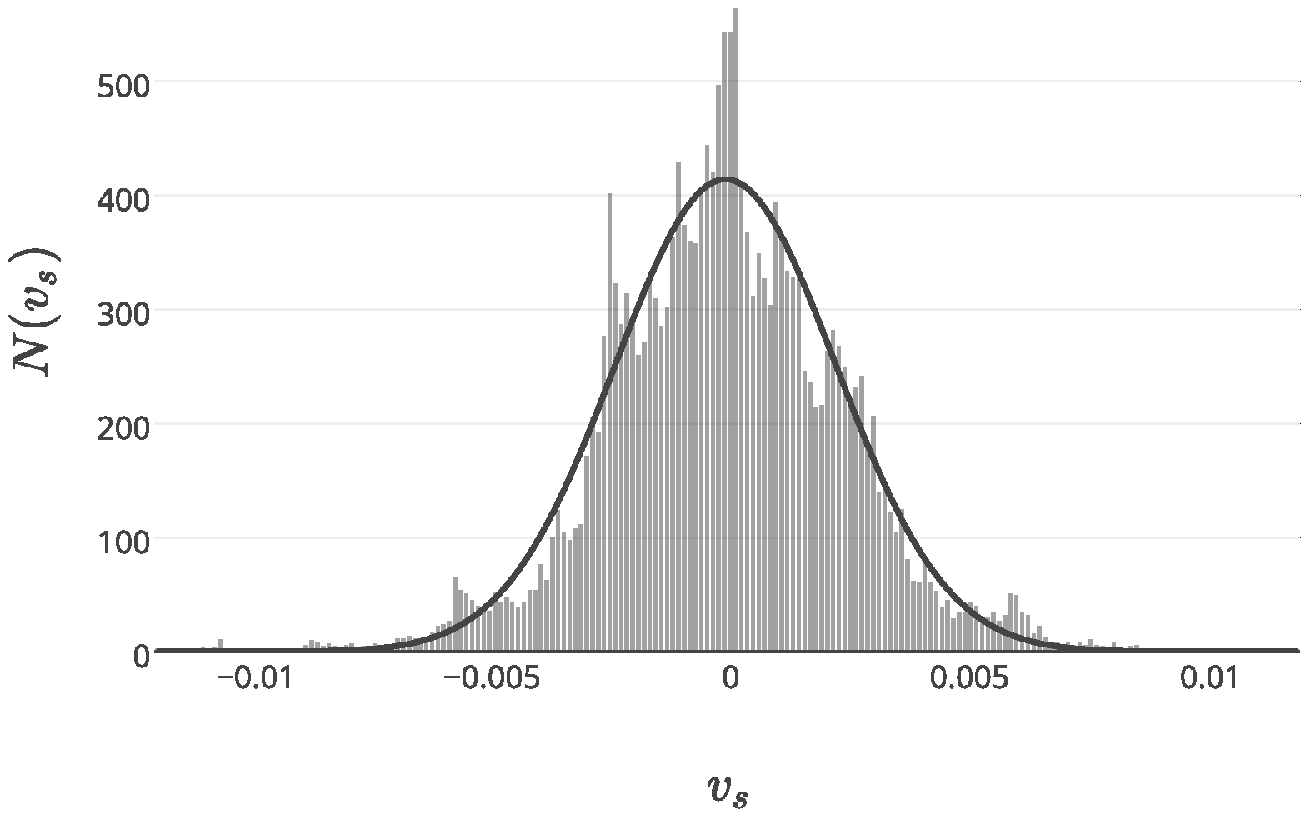}
		\caption{SHM, $\Theta = 0.001$}
	\end{subfigure}
\caption{The ion velocity distribution for the case of light ions.}
\label{fig4}
\end{figure}

For both models, the ion kinetic energy in the assumed units can be written as
\begin{equation}
\frac{1}{2}\int\limits_{ - 1/2}^{1/2} {d{\xi _{s0}}{{\left( {\frac{{d{\xi _s}}}{{d\tau }}} \right)}^2}},
\label{eq15}
\end{equation}
\noindent and the energies of collective excitations for ZHM and SHM, respectively, reduce to 
\begin{equation}
\frac{1}{{8{\pi ^2}}}\frac{m}{M}\frac{1}{{n_M^2}}\frac{\delta }{{{\omega _{pe}}}}\sum\limits_n {|{M_n}{|^2}}, \qquad
\frac{1}{{8{\pi ^2}}}\frac{m}{M}\sum\limits_n {\frac{1}{{{n^2}}}} \left[ {1 - J_0^2({a_n}) + \frac{2}{3}{J_2}^2({a_n})} \right]|{M_n}{|^2}.
\end{equation}
\noindent Note that in Zakharov's model these oscillations are referred to as ion-sound waves.

Figure \ref{fig3} demonstrates the time evolution of the ion kinetic energy and LF field energy. It should be noted that the energy of the LF field is far smaller than ion energies in all cases. Reducing the field energy  with time is caused by the energy transfer to ions as well as by the destruction of plasma density cavities, \cite{Sigov.1979}.

The rate of the HF field burnout within density cavities is determined by the value $\Theta=\theta/\delta$. It is of interest how the simulation results depend on this parameter. Obviously, the decrease of this parameter not only inhibits the burnout of the HF field in the cavities, but also broadens the spectrum of HF modes, i.e. it increases the contribution of small-scale components that leads to the deepening of plasma density cavities and to the growth of the kinetic energy of ions ejected from the cavities.

Note that for both models the ion velocity distribution function approaches the Maxwellian curve with decreasing damping rate of HF modes, as may be seen from Fig. \ref{fig4}.

Table~\ref{Tab2} demonstrates the extent of deviation of the ion velocity distribution function, obtained by numerical modeling, from the fitted Maxwellian curve for the cases shown in Fig.~(\ref{fig4}).

\begin{table}[b]
\caption{Deviation of the ion velocity distribution function, obtained by numerical simulation, from the fitted Maxwellian curve}
\begin{ruledtabular}
\begin{tabular}{ccc}
Damping rate & ZHM& SHM \\ \hline
$\Theta=0.05$   &  19.9\%   &  13\% \\
$\Theta=0.015$   &  9.9\%   &  13.4\% \\
$\Theta=0.001$   &  6.9\%   &  8.8\% \\
\end{tabular}
\end{ruledtabular}
\label{Tab2}
\end{table}

\begin{figure}[t] \centering
	\begin{subfigure}[b]{0.47\textwidth}
		\includegraphics[width=\textwidth]{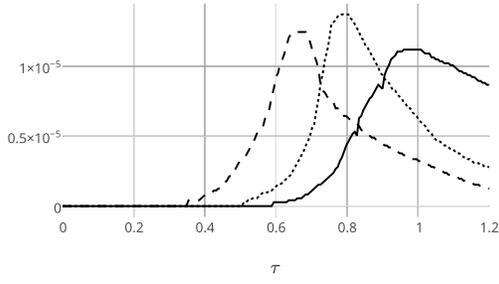}
		\caption{ZHM}
	\end{subfigure}
	\begin{subfigure}[b]{0.47\textwidth}
		\includegraphics[width=\textwidth]{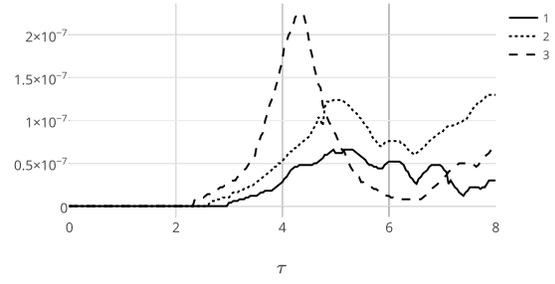}
		\caption{SHM}
	\end{subfigure}
	
\caption{Time evolution of the LF spectrum energy for the case of light ions, \\ 1~--~$\Theta = 0.05$, 2~--~$\Theta = 0.015$, 3~--~$\Theta = 0.001$.}
\label{fig5}
\end{figure}

\begin{figure}[t] \centering
	\begin{subfigure}[b]{0.47\textwidth}
		\includegraphics[width=\textwidth]{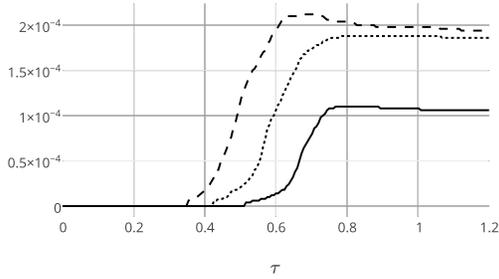}
		\caption{ZHM}
	\end{subfigure}
	\begin{subfigure}[b]{0.47\textwidth}
		\includegraphics[width=\textwidth]{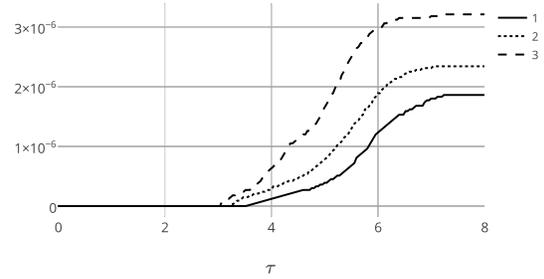}
		\caption{SHM}
	\end{subfigure}
	
\caption{Time evolution of the ion kinetic energy for the case of light ions, \\ 1 -- $\Theta=0.05$, 2 -- $\Theta=0.015$, 3 -- $\Theta=0.001$.}
\label{fig6}
\end{figure}

Figure (\ref{fig5}) shows that in the case of a non-isothermal plasma the maximum energy of ion-sound oscillations remains practically unchanged as the damping rate of HF field decreases, whereas the formation of the LF spectrum  occurs with higher rate. In a cold plasma, on the contrary, the intensity of LF oscillations grows with the decrease of the damping rate of the HF field. After that, the LF spectrum is suppressed and its energy is transferred to ions. As might be expected, the energy,  transmitted to ions, increases with the decrease of the damping rate of HF oscillations  almost in the same proportion in  both non-isothermal and cold plasmas  (see. Fig.~(\ref{fig6})).

It should be noted in conclusion that the ion-density perturbations with spatial scale smaller than the Debye radius ${r_{Di}} = {v_{Ti}}/{\omega _{pi}}$ do not contribute to the formation of LF electric fields by virtue of the screening effect. The Debye radius can be estimated from the expression \cite{Kirichok1.2014}
\begin{equation}
{r_{Di}}{k_0}/2\pi  = {R_{Di}} \propto \left\langle {\frac{{{v_i}{k_0}}}{{2\pi {\gamma _L}}}} \right\rangle \left( {\frac{\delta }{{{\omega _{pe}}}}} \right){\left( {\frac{M}{{{m_e}}}} \right)^{1/2}} =  < {v_s} > \left( {\frac{\delta }{{{\omega _{pe}}}}} \right){\left( {\frac{M}{{{m_e}}}} \right)^{1/2}}.
\end{equation}

At the stage of developed instability this value is of the order of ${R_{Di}} \le {10^{ - 3}}$ and the number of spatial spectral modes of ion density does not exceed $1/{R_{Di}}$ that is in agreement with the previous analysis.

\section{Conclusion}

The most important consequence of the instability of intense Langmuir waves in plasmas is the transfer of a portion of the field energy to ions and LF plasma density oscillations. It is reasonable to consider this problem within the framework of hybrid models, where electrons are treated as a fluid and ions are regarded as particles. The dynamics of the instability of intense long-wave oscillations in both hot and cold plasmas appears to be similar \cite{Kuklin.2013, Kuznetsov.1976}.

It was noted earlier \cite{Belkin.2013} that in the case of hot plasmas ions acquire a portion of field energy that is proportional to the ratio of the field energy to the plasma thermal energy. In the case of cold plasmas, ions get a portion of field energy that is proportional to the ratio of the instability growth rate to the plasma frequency or that is the same as to the cubic root of the ratio of electron to ion masses. The energy transferred to ions in the case of heavy ions is significantly smaller than in the case of light ions. Moreover, the amount of energy transferred to ions in the case of the cold plasma is inversely proportional to the cube root of the ion mass, and in the case of the hot plasma  the portion of energy transferred to ions decreases with growth of the ion mass faster \cite{Kirichok.2014, Kirichok1.2014}. The kinetic energy distribution of ions in the SHM is characterized by a large fraction of fast particles.

In  the case of a non-isothermic plasma (ZHM), the amplitudes of modes of the LF spectrum (ion-sound waves) are of the same order in a wide range of wave numbers. In a cold plasma (SHM), the long-wave oscillations dominate in the LF spectrum. The energy of the LF field is found to be much lower than the total kinetic energy of ions for all the cases discussed above. Reducing the energy of the LF field with time happens due to the energy transfer to ions.

The decrease of the damping rate of the HF field corresponds to the slowing of the HF field burnout in the cavities and leads to the broadening of the HF spectrum that causes the deepening of the cavities and increase of the kinetic energy of ions ejected from them. It should be noted that as the absorption rate of the HF field decreases, the ion velocity distribution function approaches the Maxwellian distribution in both models under consideration. In a cold plasma, the intensity of the long-wave LF oscillations is high and it increases with the decrease of the absorption of HF modes. It is important to note that the total energy  transferred to  ions increases as the absorption of the HF spectrum reduces.

\section*{ACKNOWLEDGMENTS}
This paper was partially supported by the grant of the State Fund for Fundamental Research (project No. $\Phi$58/175-2014). The authors thank Prof.~V.I.~Karas'  for helpful comments.

\bibliography{parametric} 

\begin{thebibliography}{51}%
\makeatletter
\providecommand \@ifxundefined [1]{%
 \@ifx{#1\undefined}
}%
\providecommand \@ifnum [1]{%
 \ifnum #1\expandafter \@firstoftwo
 \else \expandafter \@secondoftwo
 \fi
}%
\providecommand \@ifx [1]{%
 \ifx #1\expandafter \@firstoftwo
 \else \expandafter \@secondoftwo
 \fi
}%
\providecommand \natexlab [1]{#1}%
\providecommand \enquote  [1]{``#1''}%
\providecommand \bibnamefont  [1]{#1}%
\providecommand \bibfnamefont [1]{#1}%
\providecommand \citenamefont [1]{#1}%
\providecommand \href@noop [0]{\@secondoftwo}%
\providecommand \href [0]{\begingroup \@sanitize@url \@href}%
\providecommand \@href[1]{\@@startlink{#1}\@@href}%
\providecommand \@@href[1]{\endgroup#1\@@endlink}%
\providecommand \@sanitize@url [0]{\catcode `\\12\catcode `\$12\catcode
  `\&12\catcode `\#12\catcode `\^12\catcode `\_12\catcode `\%12\relax}%
\providecommand \@@startlink[1]{}%
\providecommand \@@endlink[0]{}%
\providecommand \url  [0]{\begingroup\@sanitize@url \@url }%
\providecommand \@url [1]{\endgroup\@href {#1}{\urlprefix }}%
\providecommand \urlprefix  [0]{URL }%
\providecommand \Eprint [0]{\href }%
\providecommand \doibase [0]{http://dx.doi.org/}%
\providecommand \selectlanguage [0]{\@gobble}%
\providecommand \bibinfo  [0]{\@secondoftwo}%
\providecommand \bibfield  [0]{\@secondoftwo}%
\providecommand \translation [1]{[#1]}%
\providecommand \BibitemOpen [0]{}%
\providecommand \bibitemStop [0]{}%
\providecommand \bibitemNoStop [0]{.\EOS\space}%
\providecommand \EOS [0]{\spacefactor3000\relax}%
\providecommand \BibitemShut  [1]{\csname bibitem#1\endcsname}%
\let\auto@bib@innerbib\@empty
\bibitem [{\citenamefont {Silin}\ and\ \citenamefont
  {Ruhadze}(1961)}]{Silin.1961}%
  \BibitemOpen
  \bibfield  {author} {\bibinfo {author} {\bibfnamefont {V.}~\bibnamefont
  {Silin}}\ and\ \bibinfo {author} {\bibfnamefont {A.}~\bibnamefont
  {Ruhadze}},\ }\href@noop {} {\emph {\bibinfo {title} {The electromagnetic
  properties of plasma and plasma-like media}}}\ (\bibinfo  {publisher}
  {Atomizdat},\ \bibinfo {address} {Moscow},\ \bibinfo {year}
  {1961})\BibitemShut {NoStop}%
\bibitem [{\citenamefont {Basov}\ and\ \citenamefont
  {Krohin}(1964)}]{Basov.1964}%
  \BibitemOpen
  \bibfield  {author} {\bibinfo {author} {\bibfnamefont {N.}~\bibnamefont
  {Basov}}\ and\ \bibinfo {author} {\bibfnamefont {O.}~\bibnamefont {Krohin}},\
  }\href@noop {} {\bibfield  {journal} {\bibinfo  {journal} {Zh. Exp. Teor.
  Fiz.}\ }\textbf {\bibinfo {volume} {46}},\ \bibinfo {pages} {171} (\bibinfo
  {year} {1964})}\BibitemShut {NoStop}%
\bibitem [{\citenamefont {Dawson}(1964)}]{Dawson.1964}%
  \BibitemOpen
  \bibfield  {author} {\bibinfo {author} {\bibfnamefont {J.~M.}\ \bibnamefont
  {Dawson}},\ }\href {\doibase 10.1063/1.1711346} {\bibfield  {journal}
  {\bibinfo  {journal} {Physics of Fluids}\ }\textbf {\bibinfo {volume} {7}},\
  \bibinfo {pages} {981} (\bibinfo {year} {1964})}\BibitemShut {NoStop}%
\bibitem [{\citenamefont {Pashinin}\ and\ \citenamefont
  {Prokhorov}(1971)}]{Pashinin.1971}%
  \BibitemOpen
  \bibfield  {author} {\bibinfo {author} {\bibfnamefont {P.}~\bibnamefont
  {Pashinin}}\ and\ \bibinfo {author} {\bibfnamefont {A.}~\bibnamefont
  {Prokhorov}},\ }\href@noop {} {\bibfield  {journal} {\bibinfo  {journal}
  {JETP}\ }\textbf {\bibinfo {volume} {33}},\ \bibinfo {pages} {883} (\bibinfo
  {year} {1971})}\BibitemShut {NoStop}%
\bibitem [{\citenamefont {Buts}, \citenamefont {Lebedev},\ and\ \citenamefont
  {Kurilko}(2006)}]{Buts.2006}%
  \BibitemOpen
  \bibfield  {author} {\bibinfo {author} {\bibfnamefont {V.~A.}\ \bibnamefont
  {Buts}}, \bibinfo {author} {\bibfnamefont {A.~N.}\ \bibnamefont {Lebedev}}, \
  and\ \bibinfo {author} {\bibfnamefont {V.~I.}\ \bibnamefont {Kurilko}},\
  }\href@noop {} {\emph {\bibinfo {title} {The theory of coherent radiation by
  intense electron beams}}},\ Particle acceleration and detection\ (\bibinfo
  {publisher} {Springer},\ \bibinfo {address} {Berlin},\ \bibinfo {year}
  {2006})\BibitemShut {NoStop}%
\bibitem [{\citenamefont {Fainberg}(2000)}]{Fainberg.2000}%
  \BibitemOpen
  \bibfield  {author} {\bibinfo {author} {\bibfnamefont {Y.}~\bibnamefont
  {Fainberg}},\ }\href {\doibase 10.1134/1.952858} {\bibfield  {journal}
  {\bibinfo  {journal} {Plasma Physics Reports}\ }\textbf {\bibinfo {volume}
  {26}},\ \bibinfo {pages} {335} (\bibinfo {year} {2000})}\BibitemShut
  {NoStop}%
\bibitem [{\citenamefont {Kuzelev}\ and\ \citenamefont
  {Ruhadze}(1990)}]{Kuzelev.1990}%
  \BibitemOpen
  \bibfield  {author} {\bibinfo {author} {\bibfnamefont {M.}~\bibnamefont
  {Kuzelev}}\ and\ \bibinfo {author} {\bibfnamefont {A.}~\bibnamefont
  {Ruhadze}},\ }\href@noop {} {\emph {\bibinfo {title} {Electrodynamics of
  dense electron beams in a plasma}}}\ (\bibinfo  {publisher} {Nauka},\
  \bibinfo {address} {Moscow},\ \bibinfo {year} {1990})\BibitemShut {NoStop}%
\bibitem [{\citenamefont {Shapiro}\ and\ \citenamefont
  {Shevchenko}(1976)}]{Shapiro.1976}%
  \BibitemOpen
  \bibfield  {author} {\bibinfo {author} {\bibfnamefont {V.}~\bibnamefont
  {Shapiro}}\ and\ \bibinfo {author} {\bibfnamefont {V.}~\bibnamefont
  {Shevchenko}},\ }\href {\doibase 10.1007/BF01034470} {\bibfield  {journal}
  {\bibinfo  {journal} {Radiophysics and Quantum Electronics}\ }\textbf
  {\bibinfo {volume} {19}},\ \bibinfo {pages} {543} (\bibinfo {year}
  {1976})}\BibitemShut {NoStop}%
\bibitem [{\citenamefont {Kondratenko}\ and\ \citenamefont
  {Kuklin}(1988)}]{Kondratenko.1988}%
  \BibitemOpen
  \bibfield  {author} {\bibinfo {author} {\bibfnamefont {A.}~\bibnamefont
  {Kondratenko}}\ and\ \bibinfo {author} {\bibfnamefont {V.}~\bibnamefont
  {Kuklin}},\ }\href@noop {} {\emph {\bibinfo {title} {Fundamentals of plasma
  electronics}}}\ (\bibinfo  {publisher} {Energoatomizdat},\ \bibinfo {address}
  {Moscow},\ \bibinfo {year} {1988})\BibitemShut {NoStop}%
\bibitem [{\citenamefont {Silin}(1965)}]{Silin.1965}%
  \BibitemOpen
  \bibfield  {author} {\bibinfo {author} {\bibfnamefont {V.}~\bibnamefont
  {Silin}},\ }\href@noop {} {\bibfield  {journal} {\bibinfo  {journal} {Soviet
  Physics-JETP}\ }\textbf {\bibinfo {volume} {21}},\ \bibinfo {pages} {1127}
  (\bibinfo {year} {1965})}\BibitemShut {NoStop}%
\bibitem [{\citenamefont {Zakharov}(1967{\natexlab{a}})}]{Zakharov.1967}%
  \BibitemOpen
  \bibfield  {author} {\bibinfo {author} {\bibfnamefont {V.}~\bibnamefont
  {Zakharov}},\ }\href@noop {} {\bibfield  {journal} {\bibinfo  {journal} {Sov.
  Phys. JETP}\ }\textbf {\bibinfo {volume} {24}},\ \bibinfo {pages} {455}
  (\bibinfo {year} {1967}{\natexlab{a}})}\BibitemShut {NoStop}%
\bibitem [{\citenamefont {Zakharov}(1967{\natexlab{b}})}]{Zakharov1.1967}%
  \BibitemOpen
  \bibfield  {author} {\bibinfo {author} {\bibfnamefont {V.}~\bibnamefont
  {Zakharov}},\ }\href@noop {} {\bibfield  {journal} {\bibinfo  {journal} {Sov.
  Phys. JETP}\ }\textbf {\bibinfo {volume} {24}},\ \bibinfo {pages} {740}
  (\bibinfo {year} {1967}{\natexlab{b}})}\BibitemShut {NoStop}%
\bibitem [{\citenamefont {Zakharov}(1972)}]{Zakharov.1972}%
  \BibitemOpen
  \bibfield  {author} {\bibinfo {author} {\bibfnamefont {V.}~\bibnamefont
  {Zakharov}},\ }\href@noop {} {\bibfield  {journal} {\bibinfo  {journal} {Sov.
  Phys. JETP}\ }\textbf {\bibinfo {volume} {35}},\ \bibinfo {pages} {908}
  (\bibinfo {year} {1972})}\BibitemShut {NoStop}%
\bibitem [{\citenamefont {Kruer}\ \emph {et~al.}(1970)\citenamefont {Kruer},
  \citenamefont {Kaw}, \citenamefont {Dawson},\ and\ \citenamefont
  {Oberman}}]{Kruer.1970}%
  \BibitemOpen
  \bibfield  {author} {\bibinfo {author} {\bibfnamefont {W.}~\bibnamefont
  {Kruer}}, \bibinfo {author} {\bibfnamefont {P.}~\bibnamefont {Kaw}}, \bibinfo
  {author} {\bibfnamefont {J.}~\bibnamefont {Dawson}}, \ and\ \bibinfo {author}
  {\bibfnamefont {C.}~\bibnamefont {Oberman}},\ }\href {\doibase
  10.1103/PhysRevLett.24.987} {\bibfield  {journal} {\bibinfo  {journal} {Phys.
  Rev. Lett.}\ }\textbf {\bibinfo {volume} {24}},\ \bibinfo {pages} {987}
  (\bibinfo {year} {1970})}\BibitemShut {NoStop}%
\bibitem [{\citenamefont {Aliev}\ and\ \citenamefont
  {Silin}(1965)}]{Aliev.1965}%
  \BibitemOpen
  \bibfield  {author} {\bibinfo {author} {\bibfnamefont {Y.}~\bibnamefont
  {Aliev}}\ and\ \bibinfo {author} {\bibfnamefont {V.}~\bibnamefont {Silin}},\
  }\href@noop {} {\bibfield  {journal} {\bibinfo  {journal} {Sov. Phys. JETP}\
  }\textbf {\bibinfo {volume} {21}},\ \bibinfo {pages} {601} (\bibinfo {year}
  {1965})}\BibitemShut {NoStop}%
\bibitem [{\citenamefont {Silin}\ and\ \citenamefont
  {Gorbunov}(1965)}]{Gorbunov.1965}%
  \BibitemOpen
  \bibfield  {author} {\bibinfo {author} {\bibfnamefont {V.}~\bibnamefont
  {Silin}}\ and\ \bibinfo {author} {\bibfnamefont {L.}~\bibnamefont
  {Gorbunov}},\ }\href@noop {} {\bibfield  {journal} {\bibinfo  {journal}
  {Soviet Physics-JETP}\ }\textbf {\bibinfo {volume} {49}},\ \bibinfo {pages}
  {1973} (\bibinfo {year} {1965})}\BibitemShut {NoStop}%
\bibitem [{\citenamefont {Silin}(1973{\natexlab{a}})}]{Silin.1973}%
  \BibitemOpen
  \bibfield  {author} {\bibinfo {author} {\bibfnamefont {V.}~\bibnamefont
  {Silin}},\ }\href {\doibase 10.1070/PU1973v015n06ABEH005063} {\bibfield
  {journal} {\bibinfo  {journal} {Sov. Phys. Uspekhi}\ }\textbf {\bibinfo
  {volume} {15}},\ \bibinfo {pages} {742} (\bibinfo {year}
  {1973}{\natexlab{a}})}\BibitemShut {NoStop}%
\bibitem [{\citenamefont {Kruer}(1972)}]{Kruer.1972}%
  \BibitemOpen
  \bibfield  {author} {\bibinfo {author} {\bibfnamefont {W.~L.}\ \bibnamefont
  {Kruer}},\ }\href@noop {} {\enquote {\bibinfo {title} {Heating of underdense
  plasmas by intense lasers},}\ }\bibinfo {type} {Tech. Rep.}\ (\bibinfo
  {institution} {Princeton Univ., NJ (USA). Plasma Physics Lab},\ \bibinfo
  {year} {1972})\BibitemShut {NoStop}%
\bibitem [{\citenamefont {Ivanov}\ and\ \citenamefont
  {Nikulin}(1974)}]{Ivanov.1974}%
  \BibitemOpen
  \bibfield  {author} {\bibinfo {author} {\bibfnamefont {A.}~\bibnamefont
  {Ivanov}}\ and\ \bibinfo {author} {\bibfnamefont {M.}~\bibnamefont
  {Nikulin}},\ }\href@noop {} {\bibfield  {journal} {\bibinfo  {journal} {Sov.
  Phys. JETP}\ }\textbf {\bibinfo {volume} {38}},\ \bibinfo {pages} {83}
  (\bibinfo {year} {1974})}\BibitemShut {NoStop}%
\bibitem [{\citenamefont {Kim}, \citenamefont {Stenzel},\ and\ \citenamefont
  {Wong}(1974)}]{Kim.1974}%
  \BibitemOpen
  \bibfield  {author} {\bibinfo {author} {\bibfnamefont {H.}~\bibnamefont
  {Kim}}, \bibinfo {author} {\bibfnamefont {R.}~\bibnamefont {Stenzel}}, \ and\
  \bibinfo {author} {\bibfnamefont {A.}~\bibnamefont {Wong}},\ }\href@noop {}
  {\bibfield  {journal} {\bibinfo  {journal} {Phys. Rev. Lett}\ }\textbf
  {\bibinfo {volume} {33}},\ \bibinfo {pages} {886} (\bibinfo {year}
  {1974})}\BibitemShut {NoStop}%
\bibitem [{\citenamefont {Vyacheslavov}\ \emph {et~al.}(1995)\citenamefont
  {Vyacheslavov}, \citenamefont {Burmasov}, \citenamefont {Kandaurov},
  \citenamefont {Kruglyakov}, \citenamefont {Meshkov},\ and\ \citenamefont
  {Sanin}}]{Vyacheslavov.1995}%
  \BibitemOpen
  \bibfield  {author} {\bibinfo {author} {\bibfnamefont {L.}~\bibnamefont
  {Vyacheslavov}}, \bibinfo {author} {\bibfnamefont {V.}~\bibnamefont
  {Burmasov}}, \bibinfo {author} {\bibfnamefont {I.}~\bibnamefont {Kandaurov}},
  \bibinfo {author} {\bibfnamefont {E.}~\bibnamefont {Kruglyakov}}, \bibinfo
  {author} {\bibfnamefont {O.}~\bibnamefont {Meshkov}}, \ and\ \bibinfo
  {author} {\bibfnamefont {A.}~\bibnamefont {Sanin}},\ }\href@noop {}
  {\bibfield  {journal} {\bibinfo  {journal} {Physics of Plasmas
  (1994-present)}\ }\textbf {\bibinfo {volume} {2}},\ \bibinfo {pages} {2224}
  (\bibinfo {year} {1995})}\BibitemShut {NoStop}%
\bibitem [{\citenamefont {McFarland}\ and\ \citenamefont
  {Wong}(1997)}]{Mcfarland.1997}%
  \BibitemOpen
  \bibfield  {author} {\bibinfo {author} {\bibfnamefont {M.}~\bibnamefont
  {McFarland}}\ and\ \bibinfo {author} {\bibfnamefont {A.}~\bibnamefont
  {Wong}},\ }\href@noop {} {\bibfield  {journal} {\bibinfo  {journal} {Physics
  of Plasmas (1994-present)}\ }\textbf {\bibinfo {volume} {4}},\ \bibinfo
  {pages} {945} (\bibinfo {year} {1997})}\BibitemShut {NoStop}%
\bibitem [{\citenamefont {Andreev}, \citenamefont {Silin},\ and\ \citenamefont
  {Stenchikov}(1977)}]{Andreev.1977}%
  \BibitemOpen
  \bibfield  {author} {\bibinfo {author} {\bibfnamefont {N.}~\bibnamefont
  {Andreev}}, \bibinfo {author} {\bibfnamefont {V.}~\bibnamefont {Silin}}, \
  and\ \bibinfo {author} {\bibfnamefont {G.}~\bibnamefont {Stenchikov}},\
  }\href@noop {} {\bibfield  {journal} {\bibinfo  {journal} {Sov. Plasma
  Phys.}\ }\textbf {\bibinfo {volume} {3}},\ \bibinfo {pages} {1088} (\bibinfo
  {year} {1977})}\BibitemShut {NoStop}%
\bibitem [{\citenamefont {Kovrizhnykh}(1977)}]{Kovrizhnykh.1977}%
  \BibitemOpen
  \bibfield  {author} {\bibinfo {author} {\bibfnamefont {L.}~\bibnamefont
  {Kovrizhnykh}},\ }\href@noop {} {\bibfield  {journal} {\bibinfo  {journal}
  {Sov. Plasma Phys.}\ }\textbf {\bibinfo {volume} {3}},\ \bibinfo {pages}
  {1097} (\bibinfo {year} {1977})}\BibitemShut {NoStop}%
\bibitem [{\citenamefont {Buchelnikova}\ and\ \citenamefont
  {Matochkin}(1979)}]{Buchelnikova.1979}%
  \BibitemOpen
  \bibfield  {author} {\bibinfo {author} {\bibfnamefont {N.}~\bibnamefont
  {Buchelnikova}}\ and\ \bibinfo {author} {\bibfnamefont {E.}~\bibnamefont
  {Matochkin}},\ }\href@noop {} {\bibfield  {journal} {\bibinfo  {journal} {АN
  USSR, Inst. Nuclear Phys}\ ,\ \bibinfo {pages} {20}} (\bibinfo {year}
  {1979})}\BibitemShut {NoStop}%
\bibitem [{\citenamefont {Antipov}\ \emph {et~al.}(1976)\citenamefont
  {Antipov}, \citenamefont {Nezlin}, \citenamefont {Snezhkin},\ and\
  \citenamefont {Trubnikov}}]{Antipov.1976}%
  \BibitemOpen
  \bibfield  {author} {\bibinfo {author} {\bibfnamefont {S.}~\bibnamefont
  {Antipov}}, \bibinfo {author} {\bibfnamefont {M.}~\bibnamefont {Nezlin}},
  \bibinfo {author} {\bibfnamefont {E.}~\bibnamefont {Snezhkin}}, \ and\
  \bibinfo {author} {\bibfnamefont {A.}~\bibnamefont {Trubnikov}},\ }\href@noop
  {} {\bibfield  {journal} {\bibinfo  {journal} {Sov. JETP Letters}\ }\textbf
  {\bibinfo {volume} {23}},\ \bibinfo {pages} {562} (\bibinfo {year}
  {1976})}\BibitemShut {NoStop}%
\bibitem [{\citenamefont {Sagdeev}, \citenamefont {Shapiro},\ and\
  \citenamefont {Schevchenko}(1980)}]{Sagdeev.1980}%
  \BibitemOpen
  \bibfield  {author} {\bibinfo {author} {\bibfnamefont {R.}~\bibnamefont
  {Sagdeev}}, \bibinfo {author} {\bibfnamefont {V.}~\bibnamefont {Shapiro}}, \
  and\ \bibinfo {author} {\bibfnamefont {V.}~\bibnamefont {Schevchenko}},\
  }\href@noop {} {\bibfield  {journal} {\bibinfo  {journal} {Sov. Plasma
  Phys.}\ }\textbf {\bibinfo {volume} {6}},\ \bibinfo {pages} {377} (\bibinfo
  {year} {1980})}\BibitemShut {NoStop}%
\bibitem [{\citenamefont {Wong}\ and\ \citenamefont
  {Cheung}(1984)}]{Wong.1984}%
  \BibitemOpen
  \bibfield  {author} {\bibinfo {author} {\bibfnamefont {A.}~\bibnamefont
  {Wong}}\ and\ \bibinfo {author} {\bibfnamefont {P.}~\bibnamefont {Cheung}},\
  }\href@noop {} {\bibfield  {journal} {\bibinfo  {journal} {Physical Rev.
  Lett.}\ }\textbf {\bibinfo {volume} {52}},\ \bibinfo {pages} {1222} (\bibinfo
  {year} {1984})}\BibitemShut {NoStop}%
\bibitem [{\citenamefont {Cheung}\ and\ \citenamefont
  {Wong}(1985)}]{Cheung.1985}%
  \BibitemOpen
  \bibfield  {author} {\bibinfo {author} {\bibfnamefont {P.}~\bibnamefont
  {Cheung}}\ and\ \bibinfo {author} {\bibfnamefont {A.}~\bibnamefont {Wong}},\
  }\href@noop {} {\bibfield  {journal} {\bibinfo  {journal} {Phys. of Fluids
  (1958-1988)}\ }\textbf {\bibinfo {volume} {28}},\ \bibinfo {pages} {1538}
  (\bibinfo {year} {1985})}\BibitemShut {NoStop}%
\bibitem [{\citenamefont {Karfidov}\ \emph {et~al.}(1990)\citenamefont
  {Karfidov}, \citenamefont {Rubenchik}, \citenamefont {Sergeichev},\ and\
  \citenamefont {Sychev}}]{Karfidov.1990}%
  \BibitemOpen
  \bibfield  {author} {\bibinfo {author} {\bibfnamefont {D.}~\bibnamefont
  {Karfidov}}, \bibinfo {author} {\bibfnamefont {A.}~\bibnamefont {Rubenchik}},
  \bibinfo {author} {\bibfnamefont {K.}~\bibnamefont {Sergeichev}}, \ and\
  \bibinfo {author} {\bibfnamefont {I.}~\bibnamefont {Sychev}},\ }\href@noop {}
  {\bibfield  {journal} {\bibinfo  {journal} {Sov. Phys. JETP}\ }\textbf
  {\bibinfo {volume} {98}},\ \bibinfo {pages} {1592} (\bibinfo {year}
  {1990})}\BibitemShut {NoStop}%
\bibitem [{\citenamefont {Zakharov}\ \emph {et~al.}(1989)\citenamefont
  {Zakharov}, \citenamefont {Pushkarev}, \citenamefont {Rubenchik},
  \citenamefont {Sagdeev},\ and\ \citenamefont {Shvets}}]{Zakharov.1989}%
  \BibitemOpen
  \bibfield  {author} {\bibinfo {author} {\bibfnamefont {V.}~\bibnamefont
  {Zakharov}}, \bibinfo {author} {\bibfnamefont {A.}~\bibnamefont {Pushkarev}},
  \bibinfo {author} {\bibfnamefont {A.}~\bibnamefont {Rubenchik}}, \bibinfo
  {author} {\bibfnamefont {R.}~\bibnamefont {Sagdeev}}, \ and\ \bibinfo
  {author} {\bibfnamefont {V.}~\bibnamefont {Shvets}},\ }\href@noop {}
  {\bibfield  {journal} {\bibinfo  {journal} {Sov. Phys. JETP}\ }\textbf
  {\bibinfo {volume} {96}},\ \bibinfo {pages} {591} (\bibinfo {year}
  {1989})}\BibitemShut {NoStop}%
\bibitem [{\citenamefont {Galeev}\ \emph {et~al.}(1975)\citenamefont {Galeev},
  \citenamefont {Sagdeev}, \citenamefont {Sigov}, \citenamefont {Shapiro},\
  and\ \citenamefont {Shevchenko}}]{Galeev.1975}%
  \BibitemOpen
  \bibfield  {author} {\bibinfo {author} {\bibfnamefont {A.}~\bibnamefont
  {Galeev}}, \bibinfo {author} {\bibfnamefont {R.}~\bibnamefont {Sagdeev}},
  \bibinfo {author} {\bibfnamefont {Y.}~\bibnamefont {Sigov}}, \bibinfo
  {author} {\bibfnamefont {V.}~\bibnamefont {Shapiro}}, \ and\ \bibinfo
  {author} {\bibfnamefont {V.}~\bibnamefont {Shevchenko}},\ }\href@noop {}
  {\bibfield  {journal} {\bibinfo  {journal} {Sov. J. Plasma Phys.(Engl.
  Transl.);(United States)}\ }\textbf {\bibinfo {volume} {1}} (\bibinfo {year}
  {1975})}\BibitemShut {NoStop}%
\bibitem [{\citenamefont {Sigov}\ and\ \citenamefont
  {Hodyrev}(1976)}]{Sigov.1976}%
  \BibitemOpen
  \bibfield  {author} {\bibinfo {author} {\bibfnamefont {Y.}~\bibnamefont
  {Sigov}}\ and\ \bibinfo {author} {\bibfnamefont {Y.}~\bibnamefont
  {Hodyrev}},\ }in\ \href@noop {} {\emph {\bibinfo {booktitle} {USSR Academy of
  Sciences Reports}}},\ Vol.\ \bibinfo {volume} {229}\ (\bibinfo {organization}
  {USSR Academy of Sciences Publishing},\ \bibinfo {year} {1976})\ p.\ \bibinfo
  {pages} {833}\BibitemShut {NoStop}%
\bibitem [{\citenamefont {Sigov}\ and\ \citenamefont
  {Zakharov}(1979)}]{Sigov.1979}%
  \BibitemOpen
  \bibfield  {author} {\bibinfo {author} {\bibfnamefont {Y.}~\bibnamefont
  {Sigov}}\ and\ \bibinfo {author} {\bibfnamefont {V.}~\bibnamefont
  {Zakharov}},\ }\href@noop {} {\bibfield  {journal} {\bibinfo  {journal}
  {Journal de Physique Colloques}\ }\textbf {\bibinfo {volume} {40}},\ \bibinfo
  {pages} {63} (\bibinfo {year} {1979})}\BibitemShut {NoStop}%
\bibitem [{\citenamefont {Robinson}\ and\ \citenamefont
  {De~Oliveira}(1999)}]{Robinson.1999}%
  \BibitemOpen
  \bibfield  {author} {\bibinfo {author} {\bibfnamefont {P.}~\bibnamefont
  {Robinson}}\ and\ \bibinfo {author} {\bibfnamefont {G.}~\bibnamefont
  {De~Oliveira}},\ }\href@noop {} {\bibfield  {journal} {\bibinfo  {journal}
  {Physics of Plasmas (1994-present)}\ }\textbf {\bibinfo {volume} {6}},\
  \bibinfo {pages} {3057} (\bibinfo {year} {1999})}\BibitemShut {NoStop}%
\bibitem [{\citenamefont {Kuklin}(2013)}]{Kuklin.2013}%
  \BibitemOpen
  \bibfield  {author} {\bibinfo {author} {\bibfnamefont {V.}~\bibnamefont
  {Kuklin}},\ }\href@noop {} {\bibfield  {journal} {\bibinfo  {journal} {The
  Journ. of Kharkiv Nat. Univer., Phys. Ser.: Nuclei, Particles, Fields.}\
  }\textbf {\bibinfo {volume} {1041}},\ \bibinfo {pages} {20} (\bibinfo {year}
  {2013})}\BibitemShut {NoStop}%
\bibitem [{\citenamefont {Kuznetsov}(1976)}]{Kuznetsov.1976}%
  \BibitemOpen
  \bibfield  {author} {\bibinfo {author} {\bibfnamefont {E.}~\bibnamefont
  {Kuznetsov}},\ }\href@noop {} {\bibfield  {journal} {\bibinfo  {journal}
  {Fizika Plazmy}\ }\textbf {\bibinfo {volume} {2}},\ \bibinfo {pages} {327}
  (\bibinfo {year} {1976})}\BibitemShut {NoStop}%
\bibitem [{\citenamefont {Belkin}\ \emph {et~al.}(2013)\citenamefont {Belkin},
  \citenamefont {Kirichok}, \citenamefont {Kuklin}, \citenamefont {Pryjmak},\
  and\ \citenamefont {Zagorodny}}]{Belkin.2013}%
  \BibitemOpen
  \bibfield  {author} {\bibinfo {author} {\bibfnamefont {E.}~\bibnamefont
  {Belkin}}, \bibinfo {author} {\bibfnamefont {A.}~\bibnamefont {Kirichok}},
  \bibinfo {author} {\bibfnamefont {V.}~\bibnamefont {Kuklin}}, \bibinfo
  {author} {\bibfnamefont {A.}~\bibnamefont {Pryjmak}}, \ and\ \bibinfo
  {author} {\bibfnamefont {A.}~\bibnamefont {Zagorodny}},\ }\href@noop {}
  {\bibfield  {journal} {\bibinfo  {journal} {Problems of Atomic Science and
  Technology. Series: Plasma Electronics and New Methods of Acceleration}\ ,\
  \bibinfo {pages} {260}} (\bibinfo {year} {2013})}\BibitemShut {NoStop}%
\bibitem [{\citenamefont {Kirichok}\ \emph
  {et~al.}(2014{\natexlab{a}})\citenamefont {Kirichok}, \citenamefont {Kuklin},
  \citenamefont {Pryjmak},\ and\ \citenamefont {Zagorodny}}]{Kirichok.2014}%
  \BibitemOpen
  \bibfield  {author} {\bibinfo {author} {\bibfnamefont {A.}~\bibnamefont
  {Kirichok}}, \bibinfo {author} {\bibfnamefont {V.}~\bibnamefont {Kuklin}},
  \bibinfo {author} {\bibfnamefont {A.}~\bibnamefont {Pryjmak}}, \ and\
  \bibinfo {author} {\bibfnamefont {A.}~\bibnamefont {Zagorodny}},\ }\href@noop
  {} {\bibfield  {journal} {\bibinfo  {journal} {Physical bases of
  instrumentation}\ }\textbf {\bibinfo {volume} {3}},\ \bibinfo {pages} {58}
  (\bibinfo {year} {2014}{\natexlab{a}})}\BibitemShut {NoStop}%
\bibitem [{\citenamefont {Dawson}(1962)}]{Dawson.1962}%
  \BibitemOpen
  \bibfield  {author} {\bibinfo {author} {\bibfnamefont {J.}~\bibnamefont
  {Dawson}},\ }\href@noop {} {\emph {\bibinfo {title} {Investigations of Plasma
  Instabilities in One-Dimensional Plasmas}}}\ (\bibinfo  {publisher}
  {Princeton University Plasma Physics Laboratory},\ \bibinfo {address}
  {Princeton, N.J.},\ \bibinfo {year} {1962})\BibitemShut {NoStop}%
\bibitem [{\citenamefont {Wang}\ \emph {et~al.}(1996)\citenamefont {Wang},
  \citenamefont {Payne}, \citenamefont {DuBois},\ and\ \citenamefont
  {Rose}}]{Wang.1996}%
  \BibitemOpen
  \bibfield  {author} {\bibinfo {author} {\bibfnamefont {J.}~\bibnamefont
  {Wang}}, \bibinfo {author} {\bibfnamefont {G.}~\bibnamefont {Payne}},
  \bibinfo {author} {\bibfnamefont {D.}~\bibnamefont {DuBois}}, \ and\ \bibinfo
  {author} {\bibfnamefont {H.}~\bibnamefont {Rose}},\ }\href@noop {} {\bibfield
   {journal} {\bibinfo  {journal} {Physics of Plasmas (1994-present)}\ }\textbf
  {\bibinfo {volume} {3}},\ \bibinfo {pages} {111} (\bibinfo {year}
  {1996})}\BibitemShut {NoStop}%
\bibitem [{\citenamefont {Chernousenko}, \citenamefont {Kuklin},\ and\
  \citenamefont {Panchenko}(1990)}]{Kuklin.1990}%
  \BibitemOpen
  \bibfield  {author} {\bibinfo {author} {\bibfnamefont {V.}~\bibnamefont
  {Chernousenko}}, \bibinfo {author} {\bibfnamefont {V.}~\bibnamefont
  {Kuklin}}, \ and\ \bibinfo {author} {\bibfnamefont {I.}~\bibnamefont
  {Panchenko}},\ }in\ \href@noop {} {\emph {\bibinfo {booktitle} {The
  integrability and kinetic equations for solitons}}}\ (\bibinfo  {publisher}
  {AN USSR, Inst. for Theor. Phys., Naukova Dumka},\ \bibinfo {address} {Kiev,
  Ukraine},\ \bibinfo {year} {1990})\ p.\ \bibinfo {pages} {472}\BibitemShut
  {NoStop}%
\bibitem [{\citenamefont {Clark}, \citenamefont {Payne},\ and\ \citenamefont
  {Nicholson}(1992)}]{Clark.1992}%
  \BibitemOpen
  \bibfield  {author} {\bibinfo {author} {\bibfnamefont {K.~L.}\ \bibnamefont
  {Clark}}, \bibinfo {author} {\bibfnamefont {G.~L.}\ \bibnamefont {Payne}}, \
  and\ \bibinfo {author} {\bibfnamefont {D.~R.}\ \bibnamefont {Nicholson}},\
  }\href {\doibase 10.1063/1.860269} {\bibfield  {journal} {\bibinfo  {journal}
  {Physics of Fluids B: Plasma Physics}\ }\textbf {\bibinfo {volume} {4}},\
  \bibinfo {pages} {708} (\bibinfo {year} {1992})}\BibitemShut {NoStop}%
\bibitem [{\citenamefont {Henri}\ \emph {et~al.}(2013)\citenamefont {Henri},
  \citenamefont {Califano}, \citenamefont {Briand},\ and\ \citenamefont
  {Mangeney}}]{Henri.2013}%
  \BibitemOpen
  \bibfield  {author} {\bibinfo {author} {\bibfnamefont {P.}~\bibnamefont
  {Henri}}, \bibinfo {author} {\bibfnamefont {F.}~\bibnamefont {Califano}},
  \bibinfo {author} {\bibfnamefont {C.}~\bibnamefont {Briand}}, \ and\ \bibinfo
  {author} {\bibfnamefont {A.}~\bibnamefont {Mangeney}},\ }\href@noop {}
  {\bibfield  {journal} {\bibinfo  {journal} {arXiv preprint arXiv:1301.3090}\
  } (\bibinfo {year} {2013})}\BibitemShut {NoStop}%
\bibitem [{\citenamefont {Birdsall}\ and\ \citenamefont
  {Langdon}(1985)}]{Charles.1985}%
  \BibitemOpen
  \bibfield  {author} {\bibinfo {author} {\bibfnamefont {C.~K.}\ \bibnamefont
  {Birdsall}}\ and\ \bibinfo {author} {\bibfnamefont {A.~B.}\ \bibnamefont
  {Langdon}},\ }\href@noop {} {\emph {\bibinfo {title} {Plasma physics via
  computer simulation}}}\ (\bibinfo  {publisher} {Taylor \& Francis},\ \bibinfo
  {year} {1985})\BibitemShut {NoStop}%
\bibitem [{\citenamefont {Hockney}\ and\ \citenamefont
  {Eastwood}(1988)}]{Hockney.1988}%
  \BibitemOpen
  \bibfield  {author} {\bibinfo {author} {\bibfnamefont {R.~W.}\ \bibnamefont
  {Hockney}}\ and\ \bibinfo {author} {\bibfnamefont {J.~W.}\ \bibnamefont
  {Eastwood}},\ }\href@noop {} {\emph {\bibinfo {title} {Computer simulation
  using particles}}}\ (\bibinfo  {publisher} {CRC Press},\ \bibinfo {year}
  {1988})\BibitemShut {NoStop}%
\bibitem [{\citenamefont {Silin}(1973{\natexlab{b}})}]{Silin1.1973}%
  \BibitemOpen
  \bibfield  {author} {\bibinfo {author} {\bibfnamefont {V.~P.}\ \bibnamefont
  {Silin}},\ }\href@noop {} {\emph {\bibinfo {title} {Parametric Influences of
  high-energy Radiation on Plasma.}}}\ (\bibinfo {year} {1973})\BibitemShut
  {NoStop}%
\bibitem [{\citenamefont {Dwight}(1961)}]{Dwight.1961}%
  \BibitemOpen
  \bibfield  {author} {\bibinfo {author} {\bibfnamefont {H.~B.}\ \bibnamefont
  {Dwight}},\ }\href@noop {} {\bibfield  {journal} {\bibinfo  {journal} {New
  York: The MacMillan Company, (4 ed)}\ } (\bibinfo {year} {1961})}\BibitemShut
  {NoStop}%
\bibitem [{\citenamefont {Sanbonmatsu}\ \emph
  {et~al.}(2000{\natexlab{a}})\citenamefont {Sanbonmatsu}, \citenamefont {Vu},
  \citenamefont {DuBois},\ and\ \citenamefont {Bezzerides}}]{Sanbonmatsu.2000}%
  \BibitemOpen
  \bibfield  {author} {\bibinfo {author} {\bibfnamefont {K.~Y.}\ \bibnamefont
  {Sanbonmatsu}}, \bibinfo {author} {\bibfnamefont {H.~X.}\ \bibnamefont {Vu}},
  \bibinfo {author} {\bibfnamefont {D.~F.}\ \bibnamefont {DuBois}}, \ and\
  \bibinfo {author} {\bibfnamefont {B.}~\bibnamefont {Bezzerides}},\
  }\href@noop {} {\bibfield  {journal} {\bibinfo  {journal} {Physics of Plasmas
  (1994-present)}\ }\textbf {\bibinfo {volume} {7}} (\bibinfo {year}
  {2000}{\natexlab{a}})}\BibitemShut {NoStop}%
\bibitem [{\citenamefont {Sanbonmatsu}\ \emph
  {et~al.}(2000{\natexlab{b}})\citenamefont {Sanbonmatsu}, \citenamefont {Vu},
  \citenamefont {Bezzerides},\ and\ \citenamefont
  {DuBois}}]{Sanbonmatsu1.2000}%
  \BibitemOpen
  \bibfield  {author} {\bibinfo {author} {\bibfnamefont {K.~Y.}\ \bibnamefont
  {Sanbonmatsu}}, \bibinfo {author} {\bibfnamefont {H.~X.}\ \bibnamefont {Vu}},
  \bibinfo {author} {\bibfnamefont {B.}~\bibnamefont {Bezzerides}}, \ and\
  \bibinfo {author} {\bibfnamefont {D.~F.}\ \bibnamefont {DuBois}},\ }\href
  {\doibase 10.1063/1.873991} {\bibfield  {journal} {\bibinfo  {journal}
  {Physics of Plasmas (1994-present)}\ }\textbf {\bibinfo {volume} {7}},\
  \bibinfo {pages} {1723} (\bibinfo {year} {2000}{\natexlab{b}})}\BibitemShut
  {NoStop}%
\bibitem [{\citenamefont {Kirichok}\ \emph
  {et~al.}(2014{\natexlab{b}})\citenamefont {Kirichok}, \citenamefont {Kuklin},
  \citenamefont {Pryjmak},\ and\ \citenamefont {Zagorodny}}]{Kirichok1.2014}%
  \BibitemOpen
  \bibfield  {author} {\bibinfo {author} {\bibfnamefont {A.}~\bibnamefont
  {Kirichok}}, \bibinfo {author} {\bibfnamefont {V.}~\bibnamefont {Kuklin}},
  \bibinfo {author} {\bibfnamefont {A.}~\bibnamefont {Pryjmak}}, \ and\
  \bibinfo {author} {\bibfnamefont {A.}~\bibnamefont {Zagorodny}},\ }\href@noop
  {} {\bibfield  {journal} {\bibinfo  {journal} {arXiv preprint
  arXiv:1411.3011}\ } (\bibinfo {year} {2014}{\natexlab{b}})}\BibitemShut
  {NoStop}%
\end{thebibliography}%
\end{document}